 \documentclass[sort&compress,final,5p,times,twocolumn]{elsarticle}

\usepackage{amssymb}
\usepackage{amsmath}

\journal{Physica E: Low-dimensional Systems and Nanostructures}

\begin{document}
\bibliographystyle{elsarticle-num}
\begin{frontmatter}



\title{Elastic constants of graphene: Comparison of empirical potentials and DFT calculations}


\author[b]{Irina V. Lebedeva\corref{cor}}
\ead{liv\_ira@hotmail.com}
\address[b]{Nano-Bio Spectroscopy Group and ETSF, Universidad del Pa\'is Vasco, CFM CSIC-UPV/EHU, Avenida de Tolosa 72, San Sebastian 20018, Spain
}
\cortext[cor]{Corresponding author}
\author[d]{Alexander S. Minkin}
\ead{amink@mail.ru}
\address[d]{National Research Centre ``Kurchatov Institute", Kurchatov Square 1, Moscow 123182, Russia}
\author[c]{Andrey M. Popov}
\ead{popov-isan@mail.ru}
\address[c]{Institute for Spectroscopy of Russian Academy of Sciences, Fizicheskaya Street 5, Troitsk, Moscow 108840, Russia}
\author[a]{Andrey A. Knizhnik}
\ead{kniznik@kintechlab.com}
\address[a]{Kintech Lab Ltd., 3rd Khoroshevskaya Street 12, Moscow 123298, Russia}

\begin{abstract}
The capacity of popular classical interatomic potentials to describe elastic properties of graphene is tested. The Tersoff potential, Brenner reactive bond-order potentials REBO-1990, REBO-2000, REBO-2002 and AIREBO as well as LCBOP,  PPBE-G, ReaxFF-CHO and ReaxFF-C2013 are considered. Linear and non-linear elastic response of graphene under uniaxial stretching is investigated by static energy calculations. The Young's modulus, Poisson's ratio and high-order elastic moduli are verified against the reference data available from experimental measurements and  \textit{ab initio} studies. The density functional theory calculations are performed to complement the reference data on the effective Young's modulus and Poisson's ratio at small but finite elongations. It is observed that for all the potentials considered, the elastic energy deviates remarkably from the simple quadratic dependence already at elongations of several percent. Nevertheless, LCBOP provides the results consistent with the reference data and thus realistically describes in-plane deformations of graphene. Reasonable agreement is also observed for the computationally cheap PPBE-G potential. REBO-2000, AIREBO and REBO-2002 give a strongly non-linear elastic response with a wrong sign of the third-order elastic modulus and the corresponding results are very far from the reference data. The ReaxFF potentials drastically overestimate the Poisson's ratio. Furthermore, ReaxFF-C2013  shows a number of numerical artefacts at finite elongations. The bending rigidity of graphene is also obtained by static energy calculations for large-diameter carbon nanotubes. The best agreement with the experimental and  \textit{ab initio}  data  in this case is achieved using the REBO-2000, REBO-2002 and ReaxFF potentials. Therefore, none of the considered potentials adequately describes both in-plane and out-of-plane deformations of graphene.
\end{abstract}


\end{frontmatter}


\section{Introduction}
\label{}
Atomistic simulations for large systems (up to $10^6$) and long time scales (up to 1 ms) are possible using classical interatomic potentials. However, such simulations are realistic as long as the potentials are properly fitted to physical properties relevant for each specific simulation case. In the present paper we compare the performance of popular interatomic potentials for carbon with respect to elastic properties of graphene. We consider  the Tersoff potential \cite{Tersoff1988}, Brenner reactive bond-order (REBO) potentials REBO-1990 \cite{Brenner1990}, REBO-2000 \cite{Stuart2000}, AIREBO \cite{Stuart2000} and REBO-2002 \cite{Brenner2002}, long-range carbon bond-order potential (LCBOP) \cite{Los2003},  PPBE-G \cite{Wei2011}, ReaxFF-CHO \cite{Chenoweth2008} and ReaxFF-C2013 \cite{Srinivasan2015}. 

It should be noted that REBO and ReaxFF potentials are reactive, i.e. capable of describing formation and breaking of bonds between carbon atoms. The use of such reactive potentials is necessary in simulations of processes in graphene-based systems which actually occur with changes in the bond topology, such as rupture of graphene nanoribbons \cite{Zhu2011,Sun2013}, transformation of graphene flakes into fullerenes under heat treatment \cite{Lebedeva2008, Lebedeva2012} or structural rearrangements induced by electron irradiation \cite{Santana2013,Skowron2013}, \textit{etc.} Nevertheless, a very wide set of phenomena in graphene-based systems do not involve breaking or formation of bonds and are determined only by the elastic properties of graphene. Examples, which have been investigated using classical molecular dynamics (MD) simulations, include formation of graphene folds \cite{Yamaletdinov2017} and rolling of carbon nanoscrolls from graphene nanoribbons \cite{Yamaletdinov2017,Shi2010,Wang2015}. Because of considerable bending of graphene layers  in these processes, accurate description of the bending rigidity is indispensable. Carefully reproduced bending rigidity is also necessary for atomistic studies of structure \cite{Shen2012,Furuhashi2013,Li2015} and mechanical properties \cite{Shen2012,Fang2015} of carbon nanotubes (CNTs) with graphene nanoribbons inside. Structures of stacking dislocations in graphene bilayer \cite{Popov2011} or graphene/hexagonal boron nitride heterostructure \cite{Argentero2017} are determined  by the Young's moduli of the layers. Both bending and tension are relevant for description of wrinkling of graphene membrane in nanoindentation \cite{Wang2009}, self-retracting motion of graphene layers \cite{Popov2011x}, dissipation of energy during relative motion \cite{Lebedeva2012a} and vibrations \cite{Lebedeva2011} of graphene layers, and so on. Atomistic simulations using classical potentials have been also invoked to study operation of nanoelectromechanical systems based on 
 rolling and unrolling of carbon nanoscrolls \cite{Shi2010a}, bending of graphene membrane \cite{Hwang2013} and relative motion of graphene layers \cite{Hwang2014,Kang2014,Kang2016}.

The capability of classical interatomic potentials to describe elastic properties of graphene has been addressed in a number of papers \cite{Wei2011,Berinskii2010, Sgouros2016, Memarian2015, Arroyo2004, Huang2006, Reddy2006, Shen2010, Zhao2009, WenXing2004,  Jensen2015, Cranford2011,Tersoff1992,Cao2014}. Elastic constants were obtained by static energy calculations  \cite{Memarian2015, Arroyo2004, Reddy2006} and MD simulations \cite{Wei2011, Sgouros2016, Shen2010, Zhao2009, WenXing2004, Jensen2015, Cranford2011} for strained graphene. The bending rigidity was found from calculations of elastic energies of CNTs of different radii \cite{Tersoff1992, Robertson1992, Yakobson1996, Arroyo2004, Lu2009} and bended graphene layers \cite{Shen2010, Cranford2011, Gonzalez2018}. The same quantity also determines the height of ripples in graphene and was estimated from MD simulations at finite temperatures \cite{Wei2011, Liu2009}. For the Tersoff, REBO-1990 and REBO-2002 potentials, analytical expressions were derived for the elastic constants and bending rigidity in the limit of small strains  \cite{Berinskii2010, Arroyo2004, Huang2006, Lu2009}. The results on the elastic constants, however, vary significantly depending on the approach even when similar interatomic potentials are used \cite{Berinskii2010, Arroyo2004, Huang2006, Sgouros2016, Shen2010, Zhao2009, WenXing2004, Jensen2015, Cranford2011}. While temperature \cite{ Sgouros2016, Shen2010, Zhao2009, WenXing2004, Wei2011, Jensen2015, Cranford2011, Liu2009, Gonzalez2018}, inclusion of long-range interactions \cite{Memarian2015, Shen2010, Zhao2009, WenXing2004} using  the Lennard-Jones (LJ) potential and other computational details are responsible for some deviations in the results, one of the most important reasons of the discrepancies should be looked for in the intervals of strains where the linear elastic response (linear stress-strain relationship or quadratic energy-strain relationship) is assumed. In the present paper we show that non-linear contributions can be quite important for the elastic response of graphene already at strains of several percent. Furthermore, we compare the performance of different potentials following the same computational procedure for all of them.

As the benchmark data, we use the results of \textit{ab initio} calculations and experiments. The Young's modulus of about 1 TPa was measured for a single graphene layer by nanoindentation \cite{Lee2008}. Very close results were obtained previously for graphite by inelastic x-ray scattering  \cite{Bosak2007} and ultrasonic, sonic and static tests \cite{Blakslee1970}. Indeed, no significant effect of the weak interlayer interaction on the in-plane elastic constants of graphene should be expected \cite{WenXing2004, Andres2012}. In the same experiment for graphene \cite{Lee2008}, an estimate of the third-order elastic modulus was obtained. The negative sign of this modulus is typical for majority of materials and corresponds to a decreasing stiffness upon increasing the tensile strain and an increasing stiff response upon increasing the compressive strain. The bending rigidity of graphene was estimated from the experimental data on the phonon spectrum of graphite \cite{Nicklow1972}. The out-of-plane transversal acoustic band ZA in graphite or graphene corresponds to bending of a single graphene layer. While the sound velocity of this band is zero, the coefficient of the quadratic dispersion is proportional to the bending rigidity.

Density functional theory (DFT) has been successfully used to study elastic properties of graphene \cite{Wei2009, Kalosakas2013, Cadelano2012, Kudin2001, Memarian2015, Mounet2005, Jensen2015, Lebedeva2016, Liu2007, Zhou2016, Gui2008, Shao2012, Savini2011, Andres2012, Siahlo2018, Lebedeva2012b, Wei2013, Cherian2007, Gulseren2002, Kurti1998, Sanchez1999}. The results of these studies are consistent one with each other and with the experiments, though diverse approaches have been tried. The calculations were performed in the local-density approximation (LDA) \cite{Ceperley1980} and generalized gradient approximation (GGA) using the exchange-correlation functionals of Perdew, Burke and Ernzerhof (PBE) \cite{Perdew1996} and Perdew and Wang (PW91) \cite{Perdew92}. The long-range interactions were included in some papers  \cite{Jensen2015, Lebedeva2016} using the DFT-D2 \cite{Grimme2006} and vdW-DF2 \cite{Lee2010} approaches. The bending rigidity was extracted from calculations of energies of CNTs  \cite{Kudin2001, Siahlo2018, Lebedeva2012b, Wei2013, Cherian2007, Gulseren2002, Kurti1998, Sanchez1999} and dispersion of the transversal acoustic phonon band ZA in graphene \cite{Sanchez1999}.  The elastic constants were obtained by explicit analysis of the energy-strain \cite{Cadelano2012, Kudin2001, Memarian2015, Jensen2015, Lebedeva2016} and stress-strain \cite{Wei2009, Kalosakas2013, Liu2007, Shao2012, Andres2012}  dependences or using the density functional perturbation theory (DFPT) \cite{Baroni2001} like in papers \cite{Mounet2005, Savini2011}. The non-linear elastic response was also considered and was shown to be anisotropic \cite{Kudin2001, Wei2009, Liu2007, Kalosakas2013, Zhou2016, Gui2008}. In Ref. \cite{Wei2009}, the elastic constants up to the fifth order were calculated. These data, however, were obtained for large strains of tens of percent. In the present paper we perform DFT calculations of  the elastic response of graphene in the weakly non-linear regime and use them for direct verification of the classical interatomic potentials in the same interval of strains.

The paper is organized in the following way. First we briefly describe the properties which the interatomic classical potentials were fitted to. The methodology 
 is described in section 3. In section 4, we give the results on contributions of terms of different order to the elastic energy of graphene under uniaxial streching and under bending and discuss the performance of the potentials with respect to the experimental and \textit{ab initio} data. Finally conclusions are made.

\section{Interatomic potentials for carbon}
\label{}

The Tersoff potential \cite{Tersoff1988} was one of the first potentials for carbon where the dependence of the bond order on the local environment was taken into account. It was assumed that the bond order is a monotonically decreasing function of coordination of the atoms. The effect of neighbouring atoms on the bond order is not equal. The farther the atom is from the bond considered, its effect is reduced though the proper distance and angular dependence of the bond-order term. The potential was fitted to the experimental data on the lattice constant, binding energy and bulk modulus of diamond as well as binding energy of graphite. The energies of carbon dimers and simple cubic, body-centered cubic and face-centered cubic phases were fitted to the results of DFT calculations.

The Tersoff potential works well for regular systems when two bonded atoms have the same coordination and the bond between them is just single, double or triple. However, if the atoms have different coordinations, an intermediate bond order is used. This is a completely unphysical description of the situation, which should be properly described though the formation of a bond and a radical orbital. Therefore, one of the problems of the Tersoff potential is overbinding of radicals. The second problem is related to the cases of conjugated systems like graphene, where the bonds with mixed single- and double-bond character are treated as simple double bonds. To include these non-local effects, Brenner suggested to add in the REBO-1990 potential \cite{Brenner1990} a correction to the bond-order term dependent on the coordination number of each atom in the bond and on whether the system is conjugated or not. The parameters of the potential were fitted to the experimental data on atomization energies of small hydrocarbons, binding energies and lattice constants of graphite and diamond, DFT data on the binding energies and lattice constants of simple cubic and face-centered cubic phases of carbon and results of tight-binding calculations for vacancies in diamond and graphite. Two sets of parameters were proposed. The first set is more accurate in bond lengths, while the second one in force constants. In the following we consider only the second set of parameters (Table III of Ref. \cite{Brenner1990}), which is commonly used in simulations for carbon nanostructures \cite{Lebedeva2008,Lebedeva2012}. 

For the next version of this potential, REBO-2002 \cite{Brenner2002}, it was suggested to use more flexible functions with an increased number of parameters. This allowed to include  additionally in the training set stretching force constants for diamond, graphite and small hydrocarbons. Furthermore, the term describing rotation about double carbon bonds and the correction to the angular function for atoms with a small coordination at low angles were introduced. REBO-2000 \cite{Stuart2000} is the version of the potential very close to REBO-2002. AIREBO is REBO-2000 with the torsional term for single carbon bonds and non-bonding interactions included. The latter ones are treated using the LJ potential, which is switched on for atoms at distances corresponding to the graphite spacing or if they are not likely to form a chemical bond under condition that they are not vicinal in the same molecule. It was checked that AIREBO, REBO-2000 and REBO-2002 reproduce well the $C_{11}$ and $C_{12}$ elastic constants of graphene as compared to the experimental data \cite{Stuart2000,Brenner2002}. It should be also mentioned that REBO-1990   and REBO-2002 were recently reparameterized to improve description of carbon nanostructures \cite{Sinitsa2014, Sinitsa2018}. Energies of graphene edges, barrier to vacancy migration and formation energy of carbon atomic chains were fitted. The parameters changed, however, do not affect elastic properties of periodic graphene layers and the results obtained below are valid also for these reparameterized versions.

LCBOP \cite{Los2003} is the result of another attempt to introduce non-bonding interactions into REBO. In this case switching between the short-range and long-range parts of the potential is performed at distances about 2 \AA. Though the short-range part here resembles the Brenner one, it had to be reparameterized. Furthermore, it was slightly modified to optimize elastic properties of graphite and diamond and surface properties of diamond as well as to describe properly the transformation from diamond to graphite. The Morse potential was applied for the long-range interactions. The training set included the binding energies and equilibrium distances for dimer bond, triple bond, carbon chains, graphite, diamond, simple cubic and face-centered cubic phases of carbon, force constants for triple bond, chains, graphite and diamond, formation energies of vacancies in graphite and diamond, energies of carbon clusters with and without conjugated bonds and dependence of the interlayer interaction in graphite on the interlayer distance. The reference values were taken from REBO publications \cite{Brenner1990,Brenner2002} and DFT results. Additionally the elastic constants for uniaxial compression and shear of diamond ($C_{11}$ and $C_{44}$) and graphite ($C_{11}$ and $C_{66}$), the distance between first and second bi-layer in the (231)-Pandey-reconstructed (111) surface of diamond and the energy barrier for transformation from diamond to graphite were fitted. 

In non-reactive PPBE-G potential \cite{Wei2011}, the energy is represented as a sum of four terms: Morse potential for the bond energy, harmonic angle potential, dihedral potential and 1--3 repulsion term that describes interaction of atoms separated by two bonds. The parameters of the potential were obtained by matching the forces with the ones obtained by DFT calculations using the PBE exchange-correlation functional \cite{Perdew1996}. A large number of configurations of graphene generated in MD simulations at temperatures 300, 1000 and 3000 K were considered. This adaptive force matching method is described in Refs. \cite{Omololu2008} and \cite{Omololu2011}.

The total energy in ReaxFF \cite{vanDuin2001} is also computed as a sum of various contributions: bond energy and terms related to angles, torsion, conjugation, under or overcoordination, non-bonding and Coulomb interactions. Different from non-reactive force-fields, all valence terms in ReaxFF depend on the bond order determined by the bond length and go to zero smoothly as bonds break. The original potential \cite{vanDuin2001} was fitted to the experimental data on heats of formation and geometry for a large number of hydrocarbon compounds and crystals, such as graphite, diamond, buckycall and cyclohexane, and quantum chemical data on energy curves for bond dissociation and reactions of small molecules. For ReaxFF-CHO \cite{Chenoweth2008}, the training set was supplemented by the quantum chemical data on C/H/O compounds and hydrocarbon oxidation chemistry. ReaxFF-C2013 \cite{Srinivasan2015} was developed with the aim of description of carbon condensed phases and the training set in this case included also equations of state for the volumetric expansion of diamond, in-plane and out-of-plane expansion of graphite, uniaxial expansion of diamond in the (001) direction as well as heats of formation of  mono-, di-, tri-, and quadruple vacancies and the Stone-Wales defect in graphene and heats of formation of various amorphous carbon clusters. 

As seen from this brief overview, forces, i.e. elastic properties, for graphene to different extent have been taken into account in fitting of the AIREBO, REBO-2000, REBO-2002, PPBE-G, LCBOP and ReaxFF-C2013 potentials. However, the Young's modulus, Poisson's ratio and bending rigidity were not fitted explicitly and the dependence on strain was not always considered. Therefore, additional tests are required to verify performance of the potentials with respect to the elastic moduli of graphene.

It should be noted that the computational speeds of the potentials listed above are rather different. PPBE-G has the simplest expression for the potential energy and is the fastest to compute. As the complexity of the potential grows and long-range interactions are included, the calculations become more and more heavy. According to our test calculations (see Supplementary Information), the computational time increases in the row: PPBE-G $<$ Tersoff  $<$ LCBOP $\approx$ REBO-2000  $<$ AIREBO  $<$ ReaxFF-C2013 $\lesssim$  ReaxFF-CHO. 

\section{Methods: calculation of elastic properties of graphene}
\label{}
\subsection{Young's modulus and Poisson's ratio}
\label{}
The Young's modulus and Poisson's ratio of graphene at zero temperature were obtained by atomistic calculations of the potential energy of strained graphene. The calculations with the REBO-1990 \cite{Brenner1990} and REBO-2002 \cite{Brenner2002} potentials were performed using the in-house Molecular Dynamics - kinetic Monte Carlo (MD-kMC) code \cite{MDKMC}. The calculations with the Tersoff \cite{Tersoff1988}, REBO-2000 \cite{Stuart2000}, AIREBO \cite{Stuart2000}, PPBE-G \cite{Wei2011},  LCBOP \cite{Los2003}, ReaxFF-CHO \cite{Chenoweth2008} and ReaxFF-C2013 \cite{Srinivasan2015} potentials were carried out using the Large-Scale Atomic/Molecular Massively Parallel Simulator (LAMMPS) \cite{Plimpton1995}. An input table is used in LAMMPS for the radial dependence of the 1--3 repulsion term of the PPBE-G potential. The number of points in this table was increased to 2000000 to provide a smooth dependence of the energy on the elongation. 

The calculations were performed using periodic boundary conditions. The rectangular unit cell including 4 elementary unit cells along the zigzag direction and 4 elementary unit cells along the perpendicular armchair direction was studied for most of the potentials. For AIREBO, a larger simulation cell with 16 elementary unit cells along the zigzag and armchair directions was considered. The sides of such a simulation cell exceed the doubled cutoff radius of the LJ potential (about 10 \AA). The use of larger unit cells has no effect on the results. First the equilibrium bond length $l$ was obtained (Table \ref{table:bond}). Note that the Tersoff, REBO-1990, PPBE-G and ReaxFF-CHO potentials somewhat overestimate the equilibrium bond length compared to the experimental data  \cite{Bosak2007, Trucano1975, Zhao1989, Ludsteck1972, Baskin1955}, while AIREBO underestimates the equilibrium bond length. The difference, however, does not exceed 3\% and this is not particularly important in practice. The unit cell was stretched along one of the sides by 0 -- 3\% with a step of 0.01\%. For each elongation $\epsilon$, the strain of the unit cell in the perpendicular direction that minimizes the potential energy,  $\epsilon_\perp$, was found by the golden-section search. The positions of the atoms for each size of the unit cell were optimized by the conjugate gradient method till the energy change in successive iterations was less than $10^{-12}$ eV per atom. 

\begin{table}
  \centering
    \caption{Equilibrium bond length $l$ of graphene obtained using different potentials and DFT calculations at zero temperature and the experimental data at room temperature.}
   \renewcommand{\arraystretch}{1.2}
    \resizebox{0.45\textwidth}{!}{
        \begin{tabular}{*{3}{c}}
\hline
Method  & $l$ (\AA)  & Ref. \\\hline
Tersoff \cite{Tersoff1988} & 1.4606009 &  This work \\\hline
REBO-1990 \cite{Brenner1990} & 1.4506829 &  This work \\\hline
REBO-2000 \cite{Stuart2000} & 1.4203901 &  This work \\\hline
AIREBO \cite{Stuart2000} & 1.3967508 &  This work \\\hline
REBO-2002 \cite{Brenner2002} & 1.4203901 &  This work \\\hline
LCBOP \cite{Los2003} & 1.4198886 &  This work \\\hline
PPBE-G \cite{Wei2011} & 1.4473672 &  This work \\\hline
ReaxFF-CHO \cite{Chenoweth2008} & 1.4438496  &  This work \\\hline
ReaxFF-C2013 \cite{Srinivasan2015} & 1.4215522 &  This work \\\hline
DFT (PBE) & 1.4246 & This work \\\hline
Exp. & 1.422 & \cite{Bosak2007} \\\hline
Exp. & 1.423 & \cite{Trucano1975} \\\hline
Exp. & 1.421 & \cite{Zhao1989} \\\hline
Exp. & $1.42097 \pm 0.00006$ & \cite{Ludsteck1972} \\\hline
Exp. & $1.4196 \pm 0.0003$ & \cite{Baskin1955} \\\hline
\end{tabular}
}
\label{table:bond}
\end{table}

The Poisson's ratio at each elongation  $\epsilon$ was computed as $\nu(\epsilon)=-\epsilon_\perp (\epsilon)/\epsilon$. At very small elongations below 0.1\%, the dependence of the elastic energy on the strain $\epsilon_\perp$ in the perpendicular direction is not a smooth function because of the computational noise. In such cases, the golden-section search finds one local minimum among many. As a result, bending and oscillations of the calculated  dependence of the Poisson's ratio on the elongation are observed (Fig. \ref{fig:Y}a--i). For this reason, direct calculations of the Poisson's ratio at small elongations are not reliable and to estimate its value in the limit of zero elongation, $\nu_0=\lim_{\epsilon\to0}\nu(\epsilon)$, we approximated the dependence $\nu(\epsilon)$ by a polynom in the interval of elongations from 0.1\% to 3\% (except for ReaxFF-C2013, where there is a jump discontinuity in the dependence of the Poisson's ratio on the elongation as discussed in the next section and the interval from 0.1\% to 0.95\% was used). The zeroth order term of this polynom gave us an estimate of the Poisson's ratio at zero elongation. The least-squares polynomial regression was performed using the Levenberg-Marquardt algorithm \cite{Levenberg,Marquardt}. The order of the polynom for each potential and direction of graphene stretching was chosen so that the standard error of this term was minimal. High orders of the polynom (up to 7) were required for REBO-2000, AIREBO and REBO-2002. For other potentials, the polynoms within the third order were sufficient. The values of the Poisson's ratio at zero elongation computed in this way for the zigzag and armchair directions were the same within $6\cdot10^{-5}$. A similar variation in the values was observed when a smaller interval of elongation down to 0.5\% was considered provided that the order of the polynom was properly readjusted again. This can be considered as an error of our estimates of the Poisson's ratio at zero elongation.

The common way to calculate the Young's modulus is to find an interval of elongations where the dependence of the energy on the elongation is parabolic or the stress-strain curve is linear.  For some of the potentials considered, however, the region of elongations where the elastic response is linear is very short and given the computational noise problems at small elongations, it is difficult to evaluate the Young's modulus with a satisfactory accuracy. Therefore, to estimate the Young's modulus we also approximated the dependence of the energy on the elongation by a polynom. The order of the polynom for each potential and direction of graphene stretching was chosen so that the standard error of the term corresponding to the Young's modulus was minimal. Normally polynomial fitting was performed in the range of elongations from 0.01\% to 3\%. For REBO-2000, AIREBO, REBO-2002, ReaxFF-C2013, however, even the 9-th order polynom was not sufficient. In these cases, smaller intervals of elongations down to 1\% (0.95\% for ReaxFF-C2013) were considered. It should be mentioned that the calculated values of the Young's modulus for  the zigzag and armchair directions were the same within $4\cdot10^{-5}$ TPa for all the potentials. When a smaller interval of elongations down to 0.5\% was considered and the order of the polynom was readjusted again, the estimated Young's modulus changed by no more than  $3\cdot10^{-4}$ TPa. Therefore, such an approach allowed us to get accurate estimates of the Young's modulus. Within the same approach, we could also estimate higher-order elastic moduli, such as the third-order one, $Y_3$. From the analysis of dependences of the estimated values on the maximal elongation considered, we conclude that the relative error of $Y_3$ hardly exceeds 1\%.

More exactly, the potential energy per unit volume $U(\epsilon)$ was found from the minimal potential energy per atom at each elongation $\epsilon$ as $U(\epsilon) = E(\epsilon)/(h\sigma_0)$, where the interlayer distance in graphite, $h = 3.34$ \AA, was used for the layer thickness and $\sigma_0=3\sqrt{3}l^2/4$ is the area per atom. 
The change in the potential energy per unit volume relative to the fully relaxed layer, $\Delta U(\epsilon) = U(\epsilon)-U(0)$, was approximated as
\begin{equation} \label{eq_U}
\Delta U(\epsilon) = \sum_{n=2}^{n_\mathrm{max}}  \frac{Y_n\epsilon^n}{n!}.
\end{equation}

The coefficients $Y_n$ correspond to derivatives of the energy at zero elongation
\begin{equation} \label{eq_Yn}
Y_n=\frac{d^n\Delta U}{d\epsilon^n}\Bigg|_{\epsilon=0}
\end{equation}
and $Y_2$ is just the Young's modulus.  To evaluate the importance of terms of different order to the elastic energy, relative contributions of these terms
$\zeta_n (\epsilon_\mathrm{max}) = \Delta U_n(\epsilon_\mathrm{max})/\Delta U(\epsilon_\mathrm{max})$, where $\Delta U_n = Y_n\epsilon^n/n!$, were calculated at the maximal elongation $\epsilon_\mathrm{max}$ considered for the polynomial fitting. 

The effective Young's modulus dependent on the elongation $Y(\epsilon) = d^2 \Delta U(\epsilon)/d\epsilon^2$ was found by differentiation of the fitted polynoms given by Eq. (\ref{eq_U}):
\begin{equation} \label{eq_Y}
Y(\epsilon) = \sum_{n=2}^{n_\mathrm{max}}  \frac{Y_n\epsilon^{(n-2)}}{(n-2)!}.
\end{equation}

\subsection{Bending rigidity}
\label{}
Zigzag CNTs ($100m$,0) with $m$ from 1 to 20 (the radius $R$ exceeds 3.9 nm) were considered to calculate the bending rigidity $D_2$ of graphene. Numerous papers \cite{Kudin2001, Lebedeva2012b, Wei2013, Kurti1998, Sanchez1999, Robertson1992, Arroyo2004, Lu2009} demonstrate that the bending  rigidity does not depend on the CNT chirality. The periodic boundary conditions were used. For most of the potentials, the simulation cell included 4 elementary unit cells of the CNTs along the nanotube axis. For AIREBO, 6 elementary unit cells were considered. The size of the unit cell and positions of atoms within the unit cell were optimized.

The elastic energy of a graphene layer folded into a CNT per unit area, $U_\mathrm{CNT}(R)$, was found from the potential energy of the CNT per atom, $E_\mathrm{CNT}(R)$, as $U_\mathrm{CNT}(R) = E_\mathrm{CNT}(R)/\sigma_0$. Following the approach similar to the one for evaluation of the Young's modulus, the potential energy of CNTs relative to graphene $\Delta U_\mathrm{CNT}(R) = U_\mathrm{CNT}(R)-U_\mathrm{CNT}(\infty)$, was approximated as
\begin{equation} \label{eq_UCNT}
\Delta U_\mathrm{CNT}(R)= \sum_{n=1}^{n_\mathrm{max}}  \frac{D_{2n}}{(2n)!R^{2n}}.
\end{equation}

\subsection{DFT calculations}
\label{}
To get the reference data on the dependence of the effective Young's modulus and Poisson's ratio on the elongation in the interval of 0 -- 3\% we have performed DFT calculations using the VASP code \cite{Kresse1996}. The rectangular model cell including 4 atoms and of 20~\AA~height was studied under periodic boundary conditions. It was checked that the use of a larger 16-atom cell provides identical results.
The PBE exchange-correlation functional \cite{Perdew1996} was considered. The interaction of valence electrons with atomic cores were described using the projector augmented-wave method \cite{Kresse1999}. The Monkhorst-Pack method \cite{Monkhorst1976} was applied to integrate over the Brillouin zone using 24 k-points in the armchair direction and 36 k-points in the zigzag direction. The Gaussian smearing of 0.05 eV was applied. The low convergence threshold of the self-consistent field of $10^{-10}$ eV and the high maximal energy of plane wave basis set of 2000 eV were needed to get sufficiently smooth dependences of the total energy on the strain. The geometry optimization of free and strained graphene was performed till the maximal residual force of $10^{-6}$ eV/\AA. 

First the bond length of graphene was optimized (Table \ref{table:bond}). Then one side of the cell was increased to introduce the elongation $\epsilon$ in the interval from 0.5 to 3\% with the step of 0.5\%. The dependence of the total energy on the strain along the perpendicular side $\epsilon_\perp$ was obtained  with optimization of positions of the atoms at each size of the unit cell. For each elongation $\epsilon$, the dependence of the energy on the ratio of strains $\epsilon_\perp/\epsilon$ was approximated by a parabola (within roughly  $\pm0.01$ from the minimum). The minimum of the parabola gave the Poisson's ratio $\nu(\epsilon)$. 

To get the effective Young's modulus at each elongation $\epsilon$, we considered systems with elongations $\epsilon+\Delta \epsilon$, where $\Delta \epsilon$ was changed in the  $\pm0.025$\% interval with the step of 0.005\%. The same Poisson's ratio $\nu(\epsilon)$ was assumed for all of these systems and the energy dependence on the elongation was approximated by a parabola. The second-order derivative of the parabola gave the effective Young's modulus $Y(\epsilon)$. An explicit calculation of the second-order derivative by the method of finite differences followed by averaging over the considered interval of $\Delta \epsilon$ gave the same results within 0.3\%. To estimate the Young's modulus at zero elongation, the Poisson's ratio was assumed to be the same as at the elongation of  0.005\%.

\section{Results}
\label{}
\subsection{Polynomial fit of energy dependence on elongation}
\label{}
\begin{figure*}
	\centering
	\includegraphics[width=\textwidth]{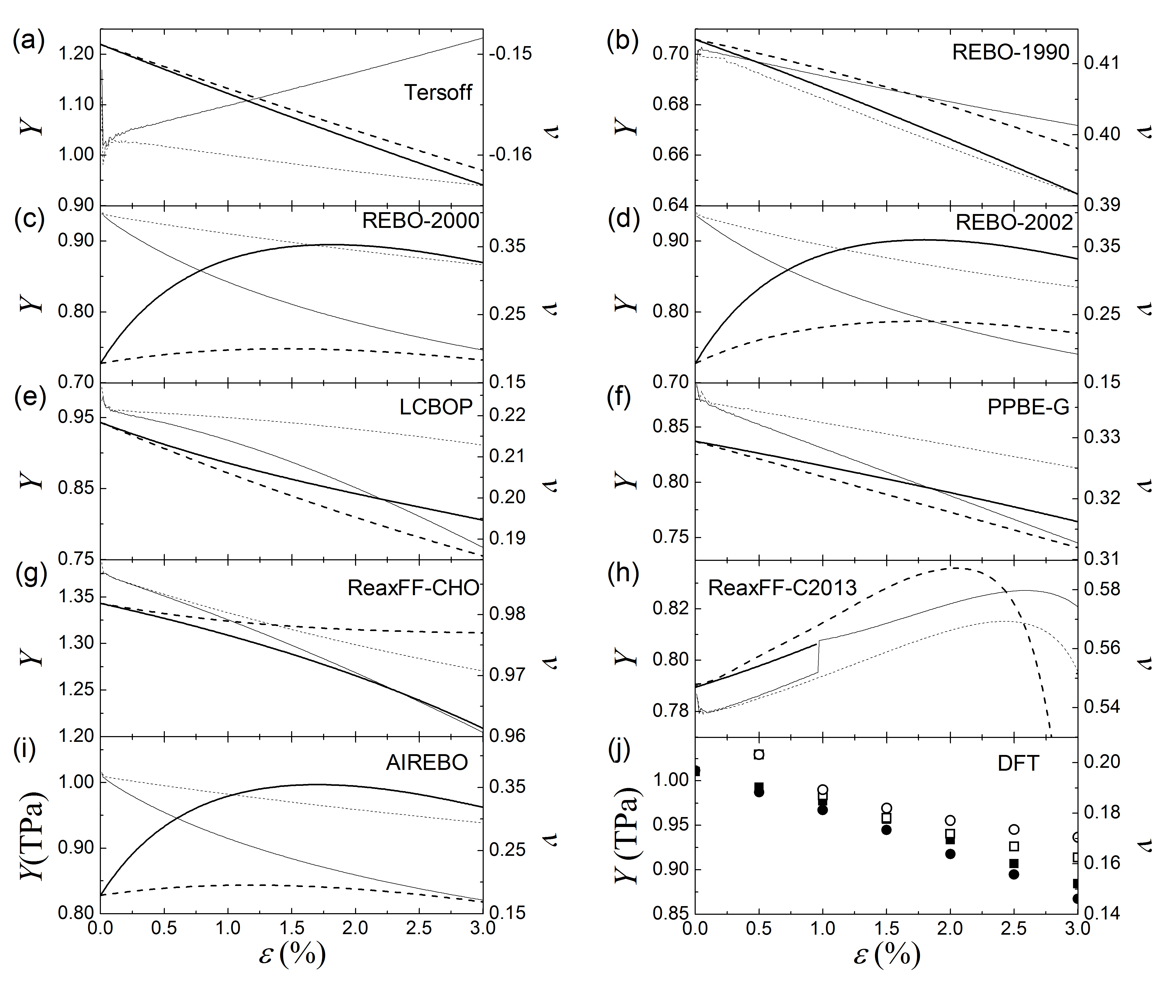}
	\caption{Effective Young's modulus $Y$ (in TPa, thick lines or filled symbols, left axis) and Poisson's ratio $\nu$ (thin lines or open symbols, right axis) of graphene under uniaxial stretching as functions of the elongation $\epsilon$ (in \%) applied in the armchair (solid lines or squares) and zigzag (dashed lines or circles) directions calculated using different approaches: (a) Tersoff, (b) REBO-1990, (c) REBO-2000, (d) REBO-2002, (e) LCBOP, (f) PPBE-G, (g) ReaxFF-CHO,  (h) ReaxFF-C2013 and (i) AIREBO potentials and (j) DFT. Note that for ReaxFF-C2013, the effective Young's modulus in the case of graphene stretching along the armchair direction is computed only up to the elongation of 0.95\%.}
	\label{fig:Y}
\end{figure*}

Let us first discuss the results of polynomial fitting of the calculated dependences of the elastic energy on the elongation upon uniaxial stretching of graphene according to Eq. \ref{eq_U}. For all the considered potentials, the dependences of the elastic energy of uniaxially stretched graphene on the elongation are non-parabolic  (Tables \ref{table:Tersoff}, \ref{table:REBO1990}, \ref{table:REBO2000},  \ref{table:AIREBO}, \ref{table:REBO2002}, \ref{table:LCBOP}, \ref{table:PPBEG}, \ref{table:REAXCHO} and \ref{table:REAXC2013}). The contributions of the third-order and further terms to the elastic energy at the maximal elongation of 3\% are of at least several percent for all the potentials. The deviation  from the simple quadratic dependence is even more evident from consideration of the effective Young's modulus (Fig. \ref{fig:Y}), which is not constant. The Poisson's ratio also changes noticeably upon increasing the elongation (Fig. \ref{fig:Y}). Note that oscillations of the Poisson's ratio at very small elongations are related to the non-smooth dependence of the elastic energy on the strain in the direction perpendicular to the elongation applied and can be ignored.

The Tersoff, REBO-1990, LCBOP, PPBE-G and ReaxFF-CHO potentials, nevertheless, provide the energy dependences relatively close to the parabolic ones and they can be fitted with  polynoms within the 7-th order. The curves obtained using the Tersoff potential  can be described according to Eq. \ref{eq_U} by polynoms of the 4-th order (Table \ref{table:Tersoff}). The 5-th order polynoms are the most appropriate for  REBO-1990, LCBOP and PPBE-G (Tables \ref{table:REBO1990}, \ref{table:LCBOP} and \ref{table:PPBEG}).  For ReaxFF-CHO, the minimal standard error of the Young's modulus $Y_2$ is reached for polynoms of the 7-th order (Table \ref{table:REAXCHO}). For these potentials, the contributions of the non-parabolic terms to the elastic energy at the maximal elongation of 3\%  do not exceed 10\% (Tables \ref{table:Tersoff}, \ref{table:REBO1990}, \ref{table:LCBOP}, \ref{table:PPBEG} and \ref{table:REAXCHO}). The effective Young's modulus and Poisson's ratio change monotonically with increasing the elongation (Fig. \ref{fig:Y}a, b and e--g) and the changes in the Poisson's ratio are within 0.03. 

For REBO-2000, REBO-2002 and AIREBO, the dependences of the elastic energy on the elongation are strongly non-parabolic. In the case of elongations in the zigzag direction, the results obtained using  REBO-2000 and AIREBO can be described by polynoms of the 6-th order (Tables \ref{table:REBO2000} and \ref{table:AIREBO}). The dependence calculated with REBO-2002 is fitted  the best by a polynom of the 8-th order  (Table \ref{table:REBO2002}). However, when the graphene layer is stretched in the armchair direction, the contributions of non-parabolic terms to the elastic energy are very large and even polynoms of the 9-th order are not sufficient. Note that even the last 9-th order terms of these polynoms provide significant contributions to the elastic energy at the maximal elongation of 3\%. In this case, a smaller interval of elongations up to 1\% is considered below to calculate the Young's modulus $Y_2$, third-order modulus $Y_3$ and Poisson's ratio $\nu_0$ at zero elongation (Tables \ref{table:REBO2000}, \ref{table:AIREBO} and \ref{table:REBO2002}). 

For REBO-2000, REBO-2002 and AIREBO, the Poisson's ratio changes drastically, from about 0.4 to about 0.2, upon streching graphene in the armchair direction by 3\% (Fig. \ref{fig:Y}c, d and i). The effective Young's modulus has a non-monotonic dependence on the elongation with a maximum at 1.5--1.8\%. It should be also noted that REBO-2000 and REBO-2002 give virtually the same results for elongations in the armchair direction  (Tables \ref{table:REBO2000} and \ref{table:REBO2002}, Fig. \ref{fig:Y}c and d). However, the data obtained for the zigzag direction are different. For example, the $Y_3$ moduli for the zigzag direction differ by a factor of 2.6. According to AIREBO, the Young's modulus is increased compared to REBO-2000 (by about 13\% at zero elongation) due to the effect of long-range interactions taken into account using the LJ potential, in agreement with Ref. \cite{Stuart2000}. However, the dependences on the strain for REBO-2000 and AIREBO are very similar (Fig. \ref{fig:Y}c and i, Tables \ref{table:REBO2000} and \ref{table:AIREBO}).

The ReaxFF-C2013 potential behaves strangely upon uniaxial stretching of graphene. There is a jump discontinuity in the dependence of the Poisson's ratio on the elongation in the armchair direction at 0.97\% (Fig. \ref{fig:Y}h). At such elongations, there are two minima in the dependence of the elastic energy on the strain $\epsilon_\perp$ in the perpendicular direction. At 0.97\%, one minimum becomes more energetically favourable than the other and a discontinuous change in the Poisson's ratio takes place. The dependences of the Poisson's ratio on the elongation for graphene stretching in the armchair and zigzag directions also have maxima at about 2.5\%. The dependence of the elastic energy on the elongation in the zigzag direction is highly non-parabolic and even the polynom of the 9-th order is not  sufficient to describe it (Table \ref{table:REAXC2013}). For the armchair direction,  the standard error of the Young's modulus is minimized for a pure parabola. However, the corresponding value of $Y_2=0.7948$ TPa differs by 0.005 TPa from the result for the zigzag direction and the results for elongations within 0.95\%. Furthermore, the parabolic dependence corresponds to the constant effective Young's modulus but this is not the case for the second-order derivative of the elastic energy evaluated numerically by the method of finite differences. Therefore, the jump discontinuity in the dependence of the Poisson's ratio  also affects the elastic energy and polynomial fitting is not appropriate for elongations including the jump. The Young's modulus and Poisson's ratio for this potential were thus computed considering only elongations within 0.95\%.

The $C_3$ rotational symmetry of graphene provides that  the Young's modulus $Y_2$ describing the linear elastic response does not depend on the direction of stretching. The moduli beyond $Y_2$, however, are anisotropic. While for the Tersoff, LCBOP, PPBE-G, ReaxFF-CHO and ReaxFF-C2013 potentials, the values of the $Y_3$ modulus for the armchair and zigzag direction differ by no more than 50\% (Tables \ref{table:Tersoff}, \ref{table:LCBOP}, \ref{table:PPBEG} and \ref{table:REAXCHO}), for the REBO potentials, this difference corresponds to a factor of 2 -- 9 (Tables \ref{table:REBO1990}, \ref{table:REBO2000}, \ref{table:AIREBO} and \ref{table:REBO2002}). Only the simplest Tersoff potential gives almost the same values for the $Y_4$ modulus in the armchair and zigzag directions (Table \ref{table:Tersoff}). For the rest of the potentials, the moduli beyond $Y_3$ are highly anisotropic. The values of the moduli beyond $Y_2$ also differ considerably between the potentials (Tables \ref{table:REBO1990}, \ref{table:REBO2000}, \ref{table:AIREBO}, \ref{table:REBO2002}, \ref{table:LCBOP}, \ref{table:PPBEG} and \ref{table:REAXCHO}).

\subsection{Young's modulus and Poisson's ratio}
\label{}
The Young's moduli $Y_2$, third-order moduli $Y_3$, Poisson's ratios $\nu_0$ in the limit of zero elongation and the corresponding literature data are summarized in Table \ref{table:Young's} (note that in some papers only elastic constants $C_{11}$, $C_{12}$, \textit{etc.} are given and $Y_2$, $Y_3$ and $\nu_0$ are evaluated from these data according to equations given in Appendix). The Young's modulus $Y_2$ and Poisson's ratio $\nu_0$ for the Tersoff potential are in agreement with previous static energy calculations \cite{Berinskii2010} and MD simulations \cite{Sgouros2016}. The Young's moduli and Poisson's ratios for the REBO potentials agree very well with the data obtained through the analytical expressions and static energy calculations at small strains \cite{Berinskii2010, Arroyo2004, Huang2006, Reddy2006}. However, for AIREBO, REBO-2000 and REBO-2002, there is some deviation from the results of MD simulations at finite temperatures \cite{Sgouros2016, Shen2010, Zhao2009, WenXing2004, Zhao2010}, which predict a higher Young's modulus and a smaller Poisson's ratio. The effect of temperature on the Young's modulus is not that strong for the second-generation REBO potentials at temperatures below 1000 K \cite{Shen2010, Zhao2010} (also according to DFT calculations \cite{Shao2012}) to explain this discrepancy. More likely, the difference in the results can be explained by large intervals of elongations where the linear elastic response was assumed in papers \cite{Sgouros2016, Shen2010, Zhao2009, WenXing2004, Zhao2010}. According to our calculations, the effective Young's modulus and Poisson's ratio for AIREBO, REBO-2000 and REBO-2002 change drastically already at such small  elongations as 0.5\%. The increased effective Young's modulus  and decreased Poisson's ratio at such elongations is consistent with the results from papers \cite{Sgouros2016, Shen2010, Zhao2009, WenXing2004, Zhao2010}. Some deviation from the MD results at room temperature \cite{Wei2011} is observed also for the PPBE-G potential. Nevertheless, the discrepancy in this case is not that significant as for the second-generation  REBO potentials and may be attributed to the effect of temperature. For the LCBOP, ReaxFF-CHO and ReaxFF-C2013 potentials, there is a good agreement with the results obtained by the MD simulations \cite{Sgouros2016, Jensen2015}.

\begin{figure}
	\centering
	\includegraphics[width=0.5\textwidth]{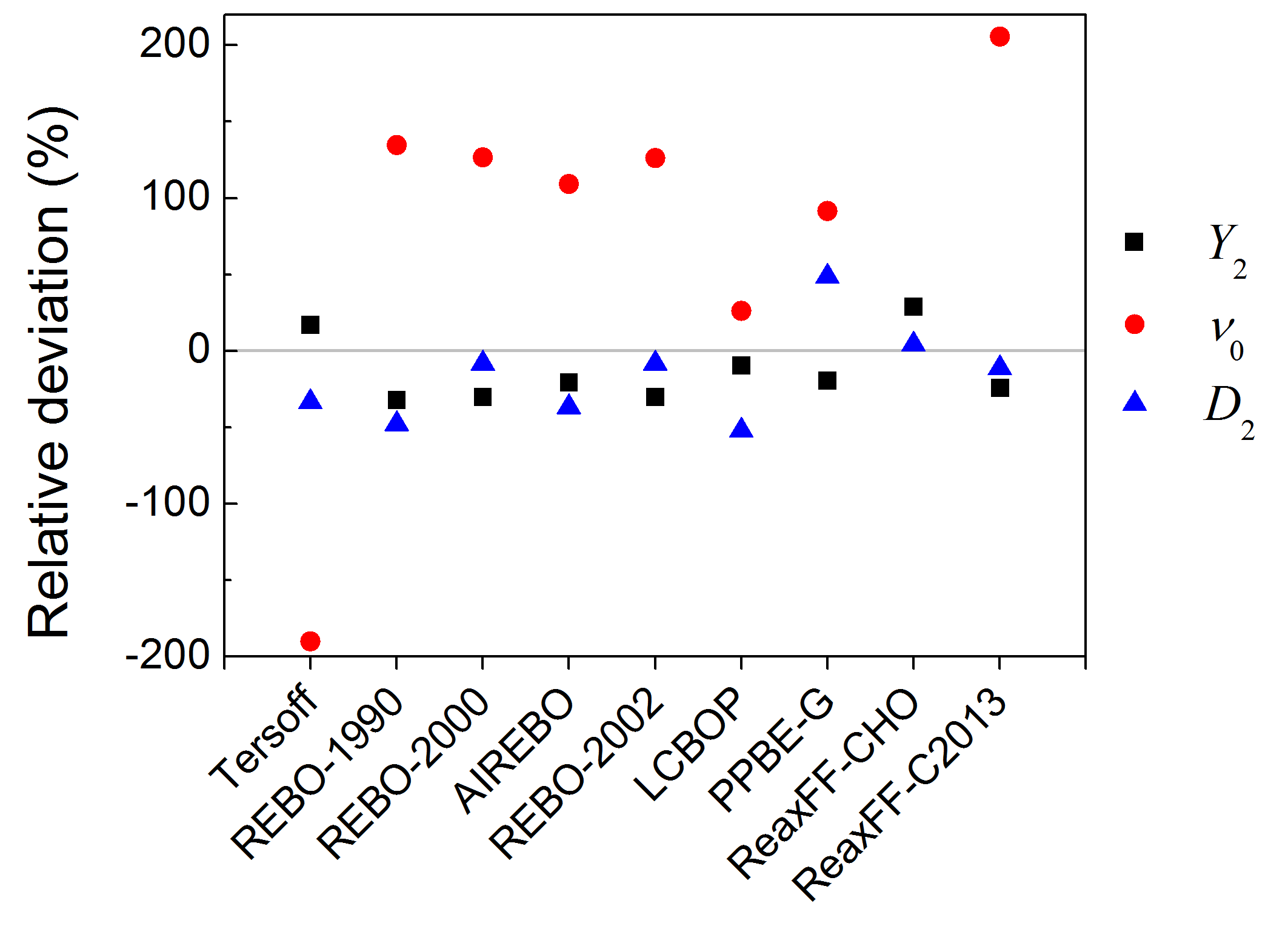}
	\caption{Relative deviation (in \%) of the Young's modulus $Y_2$ (squares), Poisson's ratio $\nu_0$ at zero elongation (circles) and bending rigidity $D_2$ (triangles) from the average reference values for different potentials. The deviation of the Poisson's ratio for ReaxFF-CHO exceeds 460\%.}
	\label{fig:dev}
\end{figure}

All the potentials considered give the Young's modulus $Y_2$ on the order 1 TPa (Table \ref{table:Young's}). To evaluate the deviations of the computed Young's moduli from the experimental \cite{Lee2008, Bosak2007, Blakslee1970} and \textit{ab initio} \cite{Wei2009, Kalosakas2013, Cadelano2012, Kudin2001, Memarian2015, Mounet2005, Jensen2015, Lebedeva2016, Liu2007, Shao2012, Savini2011, Andres2012} values, we take as a reference the average of these data, which corresponds to 1.04 TPa. The deviations of the results obtained using the interatomic potentials from this reference value are shown in  Fig. \ref{fig:dev}. It is seen the REBO potentials except AIREBO underestimate the Young's modulus by 30\%, AIREBO, ReaxFF-C2013 and PPBE-G by 20\%, while the Tersoff and ReaxFF-CHO overestimate it by 20\% and 30\%, respectively. Much more dramatic discrepancies are observed for the Poisson's ratio (Fig. \ref{fig:dev}). The experimental and \textit{ab initio} data for the Poisson's ratio range from 0.13 to 0.22 and the average reference value is 0.176. The Tersoff potential and ReaxFF-CHO are completely inadequate in this regard. The Tersoff potential gives a negative Poisson's ratio (in agreement with previous papers \cite{Sgouros2016, Berinskii2010}). The ReaxFF-CHO gives a Poisson's ratio close to unity (in agreement with \cite{Cranford2011, Jensen2015}). The Poisson's ratio is somewhat improved in the ReaxFF-C2013 but the value of about 0.5 (in agreement with \cite{Jensen2015}) is still too large. The REBO potentials also overestimate considerably the Poisson's ratio in the limit of zero elongations and give the results close to 0.4. PPBE-G somewhat overestimates the Poisson's ratio as well. Only the value for LCBOP is close to the range provided by the experimental measurements and \textit{ab initio} calculations.

Our DFT calculations for finite elongations show that the effective Young's modulus  $Y$ monotonically decreases upon stretching the graphene layer (Fig. \ref{fig:Y}j). At elongations exceeding 0.5\%, the effective Young's modulus is lower for the zigzag direction compared to the armchair one.  Among the considered potentials, only LCBOP and PPBE-G give qualitatively similar dependences (Fig. \ref{fig:Y}e and f). Quantitatively this behavior can be described using the first non-linear modulus $Y_3$ of graphene under uniaxial stretching. We roughtly estimate $Y_3$ (Table \ref{table:Young's})  based on the slope of the dependence of the effective modulus $Y$ obtained by the DFT calculations  for elongations of 1\%--3\% (Fig. \ref{fig:Y}j). 
Our estimate and previous data following from the experimental measurements \cite{Lee2008} and DFT calculations \cite{Kudin2001, Wei2009,  Kalosakas2013} (see Appendix for the relation between the elastic constants and $Y_3$ modulus) suggest that the $Y_3$ modulus is negative as can be expected for the majority of materials. AIREBO, REBO-2000 and REBO-2002, however, fail to reproduce even the sign of this modulus (Table \ref{table:Young's}). The DFT calculations \cite{Kudin2001, Wei2009, Kalosakas2013} including ours indicate that the $Y_3$ modulus for the zigzag direction is larger in magnitude than that in the armchair direction, which is equivalent to the lower elastic energy upon stretching in the zigzag direction  for the same strain if it is sufficiently small. Among the considered potentials this holds only for LCBOP and PPBE-G (Table \ref{table:Young's}). 

In the experiment \cite{Lee2008}  and Ref. \cite{Kalosakas2013}, the minimal non-linear model was used to fit the data up to large strains and the estimate of $Y_3$ of about $-4$ TPa was obtained. In Ref. \cite{Wei2009}, higher-order terms were also included in the fitting procedure and the corresponding $Y_3$ modulus was  $-(7$$-8)$ TPa. Close results of $-(6$$-9)$ TPa were obtained in Ref. \cite{Kudin2001}. Our value of about $-5$ TPa lies in the range of the previous DFT results. Nevertheless, it should be only considered as a rough estimate as we do not analyze the contribution of higher-order terms to the energy. Compared to all these data, REBO-1990 strongly underestimates the magnitude of $Y_3$ (Table \ref{table:Young's}).  PPBE-G, ReaxFF-CHO and ReaxFF-C2013 give the results close to the lower bound for $Y_3$ (Refs. \cite{Kalosakas2013,Liu2007}), while the LCBOP and Tersoff potentials agree better with the higher bound  (Refs. \cite{Wei2009, Kudin2001}). However, only the value provided by LCBOP is actually within the interval of the DFT results for $Y_3$. Considering also the relation between the $Y_3$ values for the zigzag and armchair directions, it can be concluded that LCBOP and PPBE-G describe the $Y_3$ modulus the best.

Our DFT calculations (Fig. \ref{fig:Y}j) as well as the previous ones \cite{Liu2007, Wei2009, Kalosakas2013, Zhou2016, Gui2008} indicate that the Poisson's ratio $\nu$ decreases monotonically with increasing the elongation. According to papers \cite{Liu2007, Wei2009, Zhou2016} for strains up to 30\%--60\%, the Poisson's ratio decreases almost linearly with increasing the strain and is smaller for the zigzag direction compared to the armchair one. However, in our calculations, the Poisson's ratio is almost the same for the armchair and zigzag direction at the elongation of 0.5\% (Fig. \ref{fig:Y}j) and then decreases non-linearly  taking smaller values for the armchair direction. Such a behaviour is consistent with the results obtained in papers \cite{Kalosakas2013, Gui2008} for small strains. The change of the slope in the strain dependence of the Poisson's ratio for the armchair direction was observed in paper \cite{Kalosakas2013} at the strain of about 12\%. Therefore, at small strains of up to 3\%, the Poisson's ratio should decrease faster for the armchair direction than for the zigzag one. Qualitatively similar dependences are observed at the elongations within 3\% for LCBOP, PPBE-G and ReaxFF-CHO (Fig. \ref{fig:Y}e, f and g). It should be noted, however, that none of the considered potentials describes well the shapes of the curves. 

Summarizing the results of this subsection, LCBOP gives the best results for the Young's modulus $Y_2$ and the Poisson's ratio $\nu_0$ in the limit of zero elongation. It also provides qualitatively correct dependences of the effective Young's modulus $Y$ and Poisson's ratio $\nu$ on the elongation. The corresponding values of the modulus $Y_3$ for the armchair and zigzag directions are within the range of the reported DFT results. Thus, we conclude that this potential is the most adequate for simulations of in-plane deformations in graphene among the considered interatomic potentials. Somewhat worse but also decent description of the in-plane elastic properties is provided by the computationally cheap PPBE-G potential. This can be a good choice for demanding large-scale simulations.

\begin{table*}
  \centering
    \caption{Fitted parameters $Y_n$ of the  polynomial approximation, relative standard errors $s_n$ and relative contributions $\zeta_n$ to the elastic energy of uniaxially stretched graphene for the Tersoff potential \cite{Tersoff1988} and the maximal elongation of $\epsilon_\mathrm{max}=3\%$ in the armchair and zigzag directions.}
   \renewcommand{\arraystretch}{1.2}
    \resizebox{0.7\textwidth}{!}{
        \begin{tabular}{*{7}{c}}
\hline
&\multicolumn{3}{c}{armchair} &\multicolumn{3}{c}{zigzag}  \\\hline
$n$ & $Y_n$ (TPa) & $s_n$ & $\zeta_n$ & $Y_n$ (TPa) & $s_n$ & $\zeta_n$   \\\hline
2 & 1.220  &  $2.1\cdot10^{-7}$ &  1.086 &  1.220 & $1.8\cdot10^{-7}$ & 1.076 \\\hline
3 & $-9.954$ & $7.1\cdot10^{-6}$ &  $-8.9\cdot10^{-2}$  & $-8.923$ & $6.8\cdot10^{-6}$ & $-7.9\cdot10^{-2}$\\\hline
4 & 44.50  & $1.4\cdot10^{-4}$ &  $3.0\cdot10^{-3}$ & 40.04  & $1.3\cdot10^{-4}$ & $2.7\cdot10^{-3}$  \\\hline
\end{tabular}
}
\label{table:Tersoff}
\end{table*}

\begin{table*}
  \centering
    \caption{Fitted parameters $Y_n$ of the  polynomial approximation, relative standard errors $s_n$ and relative contributions $\zeta_n$ to the elastic energy of uniaxially stretched graphene for REBO-1990 \cite{Brenner1990} and the maximal elongation of $\epsilon_\mathrm{max}=3\%$ in the armchair and zigzag directions.}
   \renewcommand{\arraystretch}{1.2}
    \resizebox{0.7\textwidth}{!}{
        \begin{tabular}{*{7}{c}}
\hline
&\multicolumn{3}{c}{armchair} &\multicolumn{3}{c}{zigzag}  \\\hline
$n$ & $Y_n$ (TPa) & $s_n$ & $\zeta_n$ & $Y_n$ (TPa) & $s_n$ & $\zeta_n$   \\\hline
2 & 0.7058 &  $6.9\cdot10^{-7}$ &  1.028 &  0.7058 & $7.1\cdot10^{-7}$ & 1.018 \\\hline
3 & $-1.815$ & $1.2\cdot10^{-4}$ &  $-2.6\cdot10^{-2}$  & $-1.036$ & $2.2\cdot10^{-4}$ & $-1.5\cdot10^{-2}$\\\hline
4 & $-18.51$  & $2.3\cdot10^{-3}$ & $-2.0\cdot10^{-3}$ & $-32.04$  & $1.4\cdot10^{-3}$ & $-3.5\cdot10^{-3}$  \\\hline
5 & $3.526\cdot10^{2}$  &  $9.7\cdot10^{-3}$ & $2.3\cdot10^{-4}$ & $5.119\cdot10^{2}$  & $6.8\cdot10^{-3}$ & $3.3\cdot10^{-4}$  \\\hline
\end{tabular}
}
\label{table:REBO1990}
\end{table*}

\begin{table*}
  \centering
    \caption{Fitted parameters $Y_n$ of the  polynomial approximation, relative standard errors $s_n$ and relative contributions $\zeta_n$ to the elastic energy of uniaxially stretched graphene for REBO-2000 \cite{Stuart2000} and the maximal elongations of $\epsilon_\mathrm{max}=3\%$ in the armchair and zigzag directions and 1\% in the armchair direction.}
   \renewcommand{\arraystretch}{1.2}
    \resizebox{\textwidth}{!}{
        \begin{tabular}{*{10}{c}}
\hline
&\multicolumn{6}{c}{$\epsilon_\mathrm{max}=3\%$} &\multicolumn{3}{c}{$\epsilon_\mathrm{max}=1\%$}  \\\hline
&\multicolumn{3}{c}{armchair} &\multicolumn{3}{c}{zigzag}  &\multicolumn{3}{c}{armchair} \\\hline
$n$ & $Y_n$ (TPa) & $s_n$ & $\zeta_n$ & $Y_n$ (TPa) & $s_n$ & $\zeta_n$  & $Y_n$ (TPa) & $s_n$ & $\zeta_n$  \\\hline
2 & 0.7273  &  $7.0\cdot10^{-6}$ &  0.858 &  0.7273 & $1.3\cdot10^{-6}$ & 0.981 & 0.7272  &  $1.8\cdot10^{-5}$ &  0.914 \\\hline
3 & 28.92  & $2.7\cdot10^{-4}$ &  0.341  & 3.133 & $2.0\cdot10^{-4}$ & $4.2\cdot10^{-2}$& 29.05  & $9.1\cdot10^{-4}$ &  0.122\\\hline
4 & $-4.593\cdot10^{3}$  & $1.4\cdot10^{-3}$ &  $-0.407$ & $-2.824\cdot10^{2}$  & $7.0\cdot10^{-4}$ & $-2.9\cdot10^{-2}$  & $-4.641\cdot10^{3}$  & $5.5\cdot10^{-3}$ & $-4.9\cdot10^{-2}$ \\\hline
5 & $7.935\cdot10^{5}$  &  $4.3\cdot10^{-3}$ & 0.421  & $1.061\cdot10^{4}$  & $3.2\cdot10^{-3}$ & $6.4\cdot10^{-3}$ & $7.454\cdot10^{5}$  &  $1.8\cdot10^{-2}$ & $1.6\cdot10^{-2}$   \\\hline
6 & $-1.469\cdot10^{8}$  &  $8.6\cdot10^{-3}$ &  $-0.390$ &$-2.959\cdot10^{5}$  & $8.6\cdot10^{-3}$ & $-9.0\cdot10^{-4}$ & $-8.702\cdot10^{7}$  &  $3.4\cdot10^{-2}$ &  $-3.0\cdot10^{-3}$\\\hline
7 & $2.334\cdot10^{10}$  &  $1.3\cdot10^{-2}$ &  0.266 &  - & - & -  &  - & - & -\\\hline
8 & $-2.596\cdot10^{12}$  &  $1.8\cdot10^{-2}$ & $-0.111$  &  - & - & - &  - & - & - \\\hline
9 & $1.458\cdot10^{14}$  &  $2.2\cdot10^{-2}$ &  $2.1\cdot10^{-2}$ & - & - & - &  - & - & -\\\hline
\end{tabular}
}
\label{table:REBO2000}
\end{table*}

\begin{table*}
  \centering
    \caption{Fitted parameters $Y_n$ of the  polynomial approximation, relative standard errors $s_n$ and relative contributions $\zeta_n$ to the elastic energy of uniaxially stretched graphene for AIREBO \cite{Stuart2000} and the maximal elongations of $\epsilon_\mathrm{max}=3\%$  in the armchair and zigzag directions and 1\% in the armchair direction.}
   \renewcommand{\arraystretch}{1.2}
    \resizebox{\textwidth}{!}{
        \begin{tabular}{*{10}{c}}
\hline
&\multicolumn{6}{c}{$\epsilon_\mathrm{max}=3\%$} &\multicolumn{3}{c}{$\epsilon_\mathrm{max}=1\%$}  \\\hline
&\multicolumn{3}{c}{armchair} &\multicolumn{3}{c}{zigzag}  &\multicolumn{3}{c}{armchair} \\\hline
$n$ & $Y_n$ (TPa) & $s_n$ & $\zeta_n$ & $Y_n$ (TPa) & $s_n$ & $\zeta_n$  & $Y_n$ (TPa) & $s_n$ & $\zeta_n$  \\\hline
2 & 0.8281  &  $8.8\cdot10^{-6}$ &  0.871 &  0.8280 & $2.4\cdot10^{-6}$ & 0.988 & 0.8280  &  $1.5\cdot10^{-5}$ &  0.921 \\\hline
3 & 30.59  & $3.6\cdot10^{-4}$ &  0.322  & 2.861 & $4.6\cdot10^{-4}$ & $3.4\cdot10^{-2}$& 30.75  & $8.4\cdot10^{-4}$ &  0.114\\\hline
4 & $-4.989\cdot10^{3}$  & $1.8\cdot10^{-3}$ &  $-0.394$ & $-3.017\cdot10^{2}$  & $1.4\cdot10^{-3}$ & $-2.7\cdot10^{-2}$  & $-5.056\cdot10^{3}$  & $4.9\cdot10^{-3}$ & $-4.7\cdot10^{-2}$ \\\hline
5 & $8.573\cdot10^{5}$  &  $5.8\cdot10^{-3}$ & 0.406  & $1.095\cdot10^{4}$  & $6.7\cdot10^{-3}$ & $5.9\cdot10^{-3}$ & $8.119\cdot10^{5}$  &  $1.6\cdot10^{-2}$ & $1.5\cdot10^{-2}$   \\\hline
6 & $-1.562\cdot10^{8}$  &  $1.2\cdot10^{-2}$ &  $-0.370$ &$-2.775\cdot10^{5}$  & $2.0\cdot10^{-2}$ & $-7.5\cdot10^{-4}$ & $-9.385\cdot10^{7}$  &  $3.1\cdot10^{-2}$ &  $-2.9\cdot10^{-3}$\\\hline
7 & $2.444\cdot10^{10}$  &  $1.8\cdot10^{-2}$ &  0.248 &  - & - & -  &  - & - & -\\\hline
8 & $-2.703\cdot10^{12}$  &  $2.5\cdot10^{-2}$ & $-0.103$  &  - & - & - &  - & - & - \\\hline
9 & $1.523\cdot10^{14}$  &  $3.1\cdot10^{-2}$ &  $1.9\cdot10^{-2}$ & - & - & - &  - & - & -\\\hline
\end{tabular}
}
\label{table:AIREBO}
\end{table*} 

\begin{table*}
  \centering
    \caption{Fitted parameters $Y_n$ of the  polynomial approximation, relative standard errors $s_n$ and relative contributions $\zeta_n$ to the elastic energy of uniaxially stretched graphene for REBO-2002 \cite{Brenner2002} and the maximal elongations of $\epsilon_\mathrm{max}=3\%$  in the armchair and zigzag directions and 1\% in the armchair direction.}
   \renewcommand{\arraystretch}{1.2}
    \resizebox{\textwidth}{!}{
        \begin{tabular}{*{10}{c}}
\hline
&\multicolumn{6}{c}{$\epsilon_\mathrm{max}=3\%$} &\multicolumn{3}{c}{$\epsilon_\mathrm{max}=1\%$}  \\\hline
&\multicolumn{3}{c}{armchair} &\multicolumn{3}{c}{zigzag}  &\multicolumn{3}{c}{armchair} \\\hline
$n$ & $Y_n$ (TPa) & $s_n$ & $\zeta_n$ & $Y_n$ (TPa) & $s_n$ & $\zeta_n$  & $Y_n$ (TPa) & $s_n$ & $\zeta_n$  \\\hline
2 & 0.7274  &  $5.0\cdot10^{-6}$ &  0.853 &  0.7273 & $8.0\cdot10^{-7}$ & 0.947 & 0.7273  &  $1.2\cdot10^{-5}$ &  0.910\\\hline
3 & 30.70  & $1.8\cdot10^{-4}$ &  0.360  & 8.529 & $8.2\cdot10^{-5}$ & 0.111 & 30.74  & $5.8\cdot10^{-4}$ &  0.128\\\hline
4 & $-5.004\cdot10^{3}$  & $9.2\cdot10^{-4}$ &  $-0.440$ & $-8.724\cdot10^{2}$  & $5.1\cdot10^{-4}$ & $-8.5\cdot10^{-2}$ & $-4.956\cdot10^{3}$  & $3.5\cdot10^{-3}$ &  $-5.2\cdot10^{-2}$  \\\hline
5 & $8.820\cdot10^{5}$  &  $2.8\cdot10^{-3}$ & 0.465  & $7.066\cdot10^{4}$  & $2.5\cdot10^{-3}$ & $4.1\cdot10^{-2}$  & $7.772\cdot10^{5}$  &  $1.1\cdot10^{-2}$ & $1.6\cdot10^{-2}$  \\\hline
6 & $-1.640\cdot10^{8}$  &  $5.6\cdot10^{-3}$ &  $-0.432$ &$-6.651\cdot10^{6}$  & $7.0\cdot10^{-3}$ & $-1.9\cdot10^{-2}$ & $-8.619\cdot10^{7}$  &  $2.3\cdot10^{-2}$ &  $-3.0\cdot10^{-3}$ \\\hline
7 & $2.593\cdot10^{10}$  &  $8.6\cdot10^{-3}$ &  0.293  & $5.600\cdot10^{8}$  & $1.3\cdot10^{-2}$ & $7.0\cdot10^{-3}$ & -  & - & -  \\\hline
8 & $-2.862\cdot10^{12}$  &  $1.2\cdot10^{-2}$ & $-0.121$  & $-2.716\cdot10^{10}$  & $1.9\cdot10^{2}$  &  $-1.3\cdot10^{-3}$ & -  & - & -\\\hline
9 & $1.594\cdot10^{14}$  &  $1.5\cdot10^{-2}$ &  $2.3\cdot10^{-2}$ & -  & - &-& -  & - & -\\\hline
\end{tabular}
}
\label{table:REBO2002}
\end{table*}

\begin{table*}
  \centering
    \caption{Fitted parameters $Y_n$ of the  polynomial approximation, relative standard errors $s_n$ and relative contributions $\zeta_n$ to the elastic energy of uniaxially stretched graphene for LCBOP \cite{Los2003} and the maximal elongation of $\epsilon_\mathrm{max}=3$\% in the armchair and zigzag directions.}
   \renewcommand{\arraystretch}{1.2}
    \resizebox{0.7\textwidth}{!}{
        \begin{tabular}{*{7}{c}}
\hline
&\multicolumn{3}{c}{armchair} &\multicolumn{3}{c}{zigzag}  \\\hline
$n$ & $Y_n$ (TPa) & $s_n$ & $\zeta_n$ & $Y_n$ (TPa) & $s_n$ & $\zeta_n$   \\\hline
2 & 0.9433 &  $1.6\cdot10^{-6}$ &  1.061 &  0.9433 & $7.2\cdot10^{-7}$ & 1.079 \\\hline
3 & $-6.689$ & $1.0\cdot10^{-4}$ &  $-7.5\cdot10^{-2}$  & $-7.586$ & $4.1\cdot10^{-5}$ & $-8.7\cdot10^{-2}$\\\hline
4 & $2.223\cdot10^{2}$  & $6.1\cdot10^{-4}$ & $1.9\cdot10^{-2}$ & $96.37$  & $6.3\cdot10^{-4}$ & $8.3\cdot10^{-3}$  \\\hline
5 & $-8.233\cdot10^{3}$  &  $1.3\cdot10^{-3}$ & $-4.2\cdot10^{-3}$ & $-8.674\cdot10^{2}$  & $5.5\cdot10^{-3}$ & $-4.5\cdot10^{-4}$  \\\hline
\end{tabular}
}
\label{table:LCBOP}
\end{table*}

\begin{table*}
  \centering
    \caption{Fitted parameters $Y_n$ of the  polynomial approximation, relative standard errors $s_n$ and relative contributions $\zeta_n$ to the elastic energy of uniaxially stretched graphene for PPBE-G \cite{Wei2011} and the maximal elongation of $\epsilon_\mathrm{max}=3\%$ in the armchair and zigzag directions.}
   \renewcommand{\arraystretch}{1.2}
    \resizebox{0.7\textwidth}{!}{
        \begin{tabular}{*{7}{c}}
\hline
&\multicolumn{3}{c}{armchair} &\multicolumn{3}{c}{zigzag}  \\\hline
$n$ & $Y_n$ (TPa) & $s_n$ & $\zeta_n$ & $Y_n$ (TPa) & $s_n$ & $\zeta_n$   \\\hline
2 & 0.8371 &  $1.7\cdot10^{-8}$ &  1.028 &  0.8371 & $1.3\cdot10^{-8}$ & 1.040 \\\hline
3 & $-2.126$ & $3.1\cdot10^{-6}$ &  $-2.6\cdot10^{-2}$  & $-3.202$ & $1.5\cdot10^{-6}$ & $-4.0\cdot10^{-2}$\\\hline
4 & $-19.39$  & $6.5\cdot10^{-5}$ & $-1.8\cdot10^{-3}$ & $-0.7871$  & $1.2\cdot10^{-3}$ & $-7.3\cdot10^{-5}$  \\\hline
5 & $49.97$  &  $2.1\cdot10^{-3}$ & $-2.8\cdot10^{-5}$ & $22.11$  & $3.4\cdot10^{-3}$ & $1.2\cdot10^{-5}$  \\\hline
\end{tabular}
}
\label{table:PPBEG}
\end{table*}

\begin{table*}
  \centering
    \caption{Fitted parameters $Y_n$ of the  polynomial approximation, relative standard errors $s_n$ and relative contributions $\zeta_n$ to the elastic energy of uniaxially stretched graphene for ReaxFF-CHO \cite{Chenoweth2008} and the maximal elongation of $\epsilon_\mathrm{max}=3\%$ in the armchair and zigzag directions.}
   \renewcommand{\arraystretch}{1.2}
    \resizebox{0.7\textwidth}{!}{
        \begin{tabular}{*{7}{c}}
\hline
&\multicolumn{3}{c}{armchair} &\multicolumn{3}{c}{zigzag}  \\\hline
$n$ & $Y_n$ (TPa) & $s_n$ & $\zeta_n$ & $Y_n$ (TPa) & $s_n$ & $\zeta_n$   \\\hline
2 & 1.343 &  $1.1\cdot10^{-7}$ &  1.028 &  1.343 & $1.5\cdot10^{-7}$ & 1.013 \\\hline
3 & $-3.035$ & $4.7\cdot10^{-5}$ &  $-2.3\cdot10^{-2}$  & $-2.500$ & $7.7\cdot10^{-5}$ & $-1.9\cdot10^{-2}$\\\hline
4 & $-74.46$  & $9.0\cdot10^{-4}$ & $-4.3\cdot10^{-3}$ & $1.253\cdot10^{2}$   & $7.2\cdot10^{-4}$ & $7.1\cdot10^{-3}$  \\\hline
5 & $4.757\cdot10^{2}$  &  $3.9\cdot10^{-2}$ & $1.6\cdot10^{-4}$ & $-1.056\cdot10^{3}$  & $2.4\cdot10^{-2}$ & $-3.6\cdot10^{-4}$  \\\hline
6 & $-5.012\cdot10^{5}$  &  $6.0\cdot10^{-3}$ & $-8.6\cdot10^{-4}$ & $-3.687\cdot10^{5}$  & $1.1\cdot10^{-2}$ & $-6.3\cdot10^{-4}$  \\\hline
7 & $2.564\cdot10^{7}$  &  $8.4\cdot10^{-3}$ & $-1.9\cdot10^{-4}$ & $2.035\cdot10^{7}$  & $1.4\cdot10^{-2}$ & $1.5\cdot10^{-4}$  \\\hline
\end{tabular}
}
\label{table:REAXCHO}
\end{table*}

\begin{table*}
  \centering
    \caption{Fitted parameters $Y_n$ of the  polynomial approximation, relative standard errors $s_n$ and relative contributions $\zeta_n$ to the elastic energy of uniaxially stretched graphene for ReaxFF-C2013 \cite{Srinivasan2015} and the maximal  elongations of $\epsilon_\mathrm{max}=3\%$ in the zigzag direction and 0.95\% in the armchair and zigzag directions.}
   \renewcommand{\arraystretch}{1.2}
    \resizebox{\textwidth}{!}{
        \begin{tabular}{*{10}{c}}
\hline
&\multicolumn{3}{c}{$\epsilon_\mathrm{max}=3\%$} &\multicolumn{6}{c}{$\epsilon_\mathrm{max}=0.95\%$} \\\hline
&\multicolumn{3}{c}{zigzag} &\multicolumn{3}{c}{armchair} &\multicolumn{3}{c}{zigzag}  \\\hline
$n$ & $Y_n$ (TPa) & $s_n$ & $\zeta_n$  & $Y_n$ (TPa) & $s_n$ & $\zeta_n$ & $Y_n$ (TPa) & $s_n$ & $\zeta_n$   \\\hline
2 & 0.7907 &  $1.4\cdot10^{-4}$ &  0.974 & 0.7894 & $5.2\cdot10^{-7}$ & 0.990 & 0.7894 & $1.5\cdot10^{-6}$ & 0.990 \\\hline
3 & $-3.513\cdot10^{-2}$ & $4.8$ &  $-4.3\cdot10^{-4}$ & 1.554 & $2.2\cdot10^{-4}$ & $6.2\cdot10^{-3}$  & $2.259$ & $4.2\cdot10^{-4}$ & $9.0\cdot10^{-3}$\\\hline
4 & $2.293\cdot10^{3}$  & $6.0\cdot10^{-2}$ & $0.212$ & 45.67 & $2.3\cdot10^{-3}$ & $4.3\cdot10^{-4}$ & $57.05$   & $4.4\cdot10^{-3}$ & $5.4\cdot10^{-4}$  \\\hline
5 & $-1.397\cdot10^{6}$  &  $5.3\cdot10^{-2}$ & $-0.774$ & -  & - & -  & -  & - & -   \\\hline
6 & $5.899\cdot10^{8}$  &  $4.6\cdot10^{-2}$ & $1.635$ & -  & - & - & -  & - & -  \\\hline
7 & $-1.657\cdot10^{11}$  &  $4.0\cdot10^{-2}$ & $-1.969$ & -  & - & - & -  & - & -   \\\hline
8 & $2.823\cdot10^{13}$  &  $3.5\cdot10^{-2}$ & $1.258$ & -  & - & -  & -  & - & -   \\\hline
9 & $-2.259\cdot10^{15}$  &  $3.1\cdot10^{-2}$ & $-0.335$ & -  & - & - & -  & - & -  \\\hline
\end{tabular}
}
\label{table:REAXC2013}
\end{table*}

\begin{table*}
  \centering
    \caption{Young's modulus$^\mathrm{a}$ $Y_2$, third-order modulus$^\mathrm{a}$ $Y_3$  and Poisson's ratio $\nu_0$ of graphene in the limit of zero elongation along the armchair/zigzag direction for different interatomic potentials, from the DFT calculations and according to the experimental data.}
   \renewcommand{\arraystretch}{1.2}
    \resizebox{0.85\textwidth}{!}{
        \begin{tabular}{*{6}{c}}
\hline
Method & Procedure & $\nu_0$ & $Y_2$ (TPa)  & $Y_3$  (TPa) &  Ref. \\\hline
Tersoff \cite{Tersoff1988} & energy-strain & $-0.158$ & 1.220 &  $-9.954$/$-8.923$ & This  work \\\hline
REBO-1990 \cite{Brenner1990}  & energy-strain & 0.412 &  0.7058  &   $-1.815$/$-1.036$ & This  work \\\hline
REBO-2000 \cite{Stuart2000}  & energy-strain & 0.398 & 0.7273 &   $29.05$/$3.133$ &  This  work\\\hline
AIREBO \cite{Stuart2000}  & energy-strain & 0.367 & 0.8280 &  30.75/2.861  &  This  work\\\hline
REBO-2002 \cite{Brenner2002}  & energy-strain & 0.397 & 0.7273  &   $30.74$/$8.529$ &  This  work\\\hline
LCBOP \cite{Los2003} & energy-strain & 0.221 &0.9433 &   $-6.689$/$-7.586$ &  This  work \\\hline
PPBE-G \cite{Wei2011} & energy-strain & 0.336 & 0.8371 &    $-2.126$/$-3.202$ &  This  work\\\hline
ReaxFF-CHO \cite{Chenoweth2008} &energy-strain& 0.987   &  1.343  & $-3.035$/$-2.500$ & This  work\\\hline
ReaxFF-C2013 \cite{Srinivasan2015} & energy-strain& 0.537 & 0.7948/0.7894  &   - /$2.271$ &  This  work \\\hline
Tersoff & analytical limit & -0.158 & 1.22 & - &   \cite{Berinskii2010}  \\\hline
Tersoff & MD at 1 K & -0.1 & 1.26 & - &    \cite{Sgouros2016}  \\\hline
Tersoff & energy-strain & - & 1.000/1.098 & - &    \cite{Memarian2015}  \\\hline
REBO-1990 & analytical limit  & 0.412 & 0.707 & - &   \cite{Berinskii2010}  \\\hline
REBO-1990 &  \begin{tabular}{@{}c@{}}  analytical limit\\  \& energy-strain \end{tabular} & 0.412 & 0.707 & - &    \cite{Arroyo2004}  \\\hline
REBO-1990 & analytical limit  & 0.412 & 0.705 & - &   \cite{Huang2006}  \\\hline
REBO-1990 &energy-strain &  0.416 & 0.681 &- &  \cite{Reddy2006}  \\\hline
REBO-2000 & MD at 1 K & 0.3 & 0.86 & - &    \cite{Sgouros2016}  \\\hline
AIREBO  & MD$^\mathrm{b}$ at 300 K & 0.197--0.223 & 0.91--0.96 & - &  \cite{Shen2010}  \\\hline
AIREBO   & MD at 300 K& $0.21\pm0.01$ & $1.01\pm0.03$ & -&  \cite{Zhao2009} \\\hline
AIREBO & energy-strain  & -  & 0.844/1.013 & - &    \cite{Memarian2015}  \\\hline
REBO-2002 &analytical limit  & 0.397 & 0.728 & -&   \cite{Berinskii2010}  \\\hline
REBO-2002 &analytical limit &  0.397 & 0.728 & -&    \cite{Arroyo2004}  \\\hline
REBO-2002 & analytical limit & 0.397 & 0.727 & -&  \cite{Huang2006}  \\\hline
REBO-2002+LJ & MD at 300 K& - & 1.049 & - & \cite{WenXing2004}  \\\hline
LCBOP & MD at 1 K & 0.2 & 0.93 & -&  \cite{Sgouros2016}  \\\hline
PPBE-G  & MD at 300 K  & $0.28\pm0.03$ & $0.95\pm0.02$ &   - &   \cite{Wei2011} \\\hline
ReaxFF-CHO & MD at 300 K  & 0.876 &  1.257  & - &  \cite{Jensen2015} \\\hline
ReaxFF-CHO  & MD at 10 K  & 1.16 & 0.971 & - & \cite{Cranford2011}  \\\hline
ReaxFF-C2013 & MD at 300 K  & 0.502 &  0.765 & - &   \cite{Jensen2015} \\\hline
PBE & energy-strain & 0.203$^\mathrm{c}$ &  1.01 & $-(4.7\pm0.1)$/$-(5.0\pm0.1)$ &  This work \\\hline
PBE & stress-strain& 0.169 &  1.042 & $-7.41$/$-8.00$ &  \cite{Wei2009} \\\hline
PBE & stress-strain& 0.22 &  0.958 & $-4.0$/$-4.2$ &   \cite{Kalosakas2013} \\\hline
PBE & energy-strain & 0.149 &  1.033 &  $-6.1$/$-8.7$  &  \cite{Kudin2001} \\\hline
PBE & energy-strain & $0.15\pm0.03$ &  $1.05\pm0.03$ &  &  \cite{Cadelano2012} \\\hline
PBE & energy-strain & & 1.104 & &   \cite{Memarian2015}  \\\hline
PBE & DFPT$^\mathrm{d}$ & 0.201 &  1.035 & - &   \cite{Mounet2005} \\\hline
PW91 & energy-strain & 0.173 & & & \cite{Gui2008}  \\\hline
PBE-D2& energy-strain & 0.139 &  1.046 & - &   \cite{Jensen2015} \\\hline
vdW-DF2 & energy-strain& 0.174 & 0.991  &  - &   \cite{Lebedeva2016} \\\hline
LDA & stress-strain & 0.186 & 1.050 & - &  \cite{Liu2007} \\\hline
LDA & stress-strain& 0.184 & 1.044 & - &  \cite{Shao2012} \\\hline
LDA & \begin{tabular}{@{}c@{}}  DFPT$^\mathrm{d}$\\  \& energy-strain \end{tabular} & 0.158 & 1.081 & - &   \cite{Savini2011} \\\hline
LDA & stress-strain & 0.191 & 1.030 & - &   \cite{Andres2012} \\\hline
LDA & DFPT$^\mathrm{d}$ & 0.210 & 1.069 & - &   \cite{Mounet2005} \\\hline
LDA & energy-strain & - & 1.099 & - &   \cite{Memarian2015}  \\\hline
Exp. & \begin{tabular}{@{}c@{}}  inelastic x-ray\\ scattering$^\mathrm{d}$ \end{tabular} &  $0.13 \pm 0.03$ &  $1.09 \pm 0.03$  & - &  \cite{Bosak2007} \\\hline
Exp.  & \begin{tabular}{@{}c@{}}  vibrational \\ \& static tests$^\mathrm{d}$ \end{tabular} & $0.17 \pm 0.02$ & $1.03 \pm 0.03$  &  - &  \cite{Blakslee1970}\\\hline
Exp.  & nanoindentation  & -  & $1.0$ & $-4.0$ & \cite{Lee2008}\\\hline
\multicolumn{6}{l}{$^\mathrm{a}$It is assumed that the thickness of a single graphene layer is $h = 3.34$ \AA~(also for the data from literature).}\\
\multicolumn{6}{l}{$^\mathrm{b}$For finite graphene flakes.}\\
\multicolumn{6}{l}{$^\mathrm{c}$For the elongation of 0.5\%.}\\
\multicolumn{6}{l}{$^\mathrm{d}$For graphite.}
\end{tabular}
}
\label{table:Young's}
\end{table*}

\subsection{Bending rigidity}
\label{}
For most of the potentials considered, the energies of ($100m$,0) CNTs, where $m$ is integer from 1 to 20, can be approximated by the 4-th order polynom of the curvature ($n_\mathrm{max} = 2$ in Eq. \ref{eq_UCNT}, Table \ref{table:bend}). Though the contribution of the 4-th order term is very small, it is included to minimize the standard error for the bending rigidity $D_2$. For REBO-2002 and PPBE-G, polynoms of the 10-th and 8-th order, respectively,  ($n_\mathrm{max} = 5$ and $n_\mathrm{max} = 4$, respectively) are needed to minimize the error for $D_2$ (Table \ref{table:bend1}). Nevertheless, the contribution of the fourth-order and further terms to the energy is very small and the values of $D_2$ almost do not change upon increasing the order of the polynom (Tables \ref{table:bend} and \ref{table:bend1}). 

The calculated values of the bending rigidity $D_2$ for the Tersoff, REBO-1990, REBO-2002 and ReaxFF-C2013 potentials (Tables \ref{table:bend}) agree with the previous static energy calculations \cite{Tersoff1992, Robertson1992, Yakobson1996, Arroyo2004, Lu2009}  and the analytical limit \cite{Arroyo2004, Huang2006, Lu2009} for small curvatures (Table \ref{table:lit}). However, our results for REBO-2002 and PPBE-G are considerably greater than the values deduced for these potentials from the height of ripples in graphene at room temperature \cite{Wei2011, Liu2009}. Our static calculations correspond to the limit of zero temperature and such a difference with the MD results is consistent with a rapidly decreasing behaviour of the bending rigidity upon increasing temperature \cite{Liu2009}. Simulations of ripples in graphene at low temperatures are required to confirm this hypothesis. The value obtained previously for ReaxFF-CHO by bending a graphene layer \cite{Cranford2011} is greater than the one from our calculations for CNTs. In that paper the graphene layer was not relaxed along the bending axis and the elastic energy dependence on the applied initial curvature was considered instead of the distribution of curvatures resulting from the geometry optimization. This explains why the bending rigidity in Ref. \cite{Cranford2011} could be overestimated.

The DFT data for the energies of CNTs correspond to the bending rigidity of 1.4--1.7 eV \cite{Kudin2001, Siahlo2018, Lebedeva2012b, Wei2013, Cherian2007, Gulseren2002, Kurti1998, Sanchez1999} (Table \ref{table:lit}). The value of 1.7 eV was obtained from the dispersion of the ZA phonon mode for graphene \cite{Sanchez1999}. The value of 1.2 eV was estimated from the phonon spectrum of graphite \cite{Nicklow1972}. We use the average value from the DFT and experimental studies of 1.53 eV as a reference. According to our calculations, the Tersoff, REBO-1990 and LCBOP potentials considerably underestimate the bending rigidity (Fig. \ref{fig:dev}). The bending rigidity for the REBO-2000, REBO-2002 and ReaxFF potentials fits well into the range of the DFT data. Such an improvement in the description of graphene bending  in the second-generation REBO potentials was attributed to the account of effects of dihedral angles \cite{Lu2009}. 

Though torsional terms are also present in the PPBE-G potential, the bending rigidity evaluated from the elastic energy of CNTs for this potential is too high. The training set for this potential included only the structures generated at room temperature and above. This may explain the good performance of PPBE-G for ripples in graphene at room temperature but poor performance 
 in the static energy calculations for CNTs. The account of torsional terms for single bonds  in AIREBO (as compared to torsional terms for double bonds only in REBO-2000 and REBO-2002)  also leads to a decrease in the bending rigidity down to 1 eV (Table \ref{table:bend}).  It should be noted, however, that AIREBO reproduces rather well elastic energies of fullerenes \cite{Wei2013}. REBO-1990 also gives better results for small fullerenes than REBO-2002 \cite{Sinitsa2014}. This can be achieved by the proper balance of the contributions coming from the mean and Gaussian curvatures \cite{Wei2013} and determined by the bending rigidity and Gaussian bending stiffness, respectively.

\begin{table*}
  \centering
    \caption{Fitted parameters $D_{2n}$ of the polynomial approximation of the 4-th order ($n_\mathrm{max} = 2$) for zigzag CNTs and relative standard errors $d_{2n}$  for different interatomic potentials.}
   \renewcommand{\arraystretch}{1.2}
    \resizebox{0.6\textwidth}{!}{
        \begin{tabular}{*{5}{c}}
\hline
Potential & $D_2$ (eV) & $d_2$ & $D_4$ (eV$\cdot$\AA$^2$) & $d_4$ \\\hline
Tersoff \cite{Tersoff1988} & 1.015 &  $1.5\cdot10^{-7}$ &  28.82 & $1.1\cdot10^{-4}$ \\\hline
REBO-1990 \cite{Brenner1990}  & 0.7970 &  $6.5\cdot10^{-8}$ &  6.810 & $1.5\cdot10^{-4}$ \\\hline
REBO-2000 \cite{Stuart2000}  & 1.402 &  $1.7\cdot10^{-7}$ &  $-11.66$ & $3.8\cdot10^{-4}$ \\\hline
AIREBO \cite{Stuart2000}  & 0.9667 &  $6.9\cdot10^{-7}$ &  $-0.5951$ & $2.0\cdot10^{-2}$ \\\hline
REBO-2002 \cite{Brenner2002}  & 1.402 &  $5.8\cdot10^{-6}$ &  $-15.88$ & $9.4\cdot10^{-3}$ \\\hline
LCBOP \cite{Los2003} & 0.7309 &  $1.7\cdot10^{-6}$ &  6.707 & $3.4\cdot10^{-3}$ \\\hline
PPBE-G \cite{Wei2011} & 2.271 &  $9.7\cdot10^{-6}$ &  $-23.82$ & $1.8\cdot10^{-2}$ \\\hline
ReaxFF-CHO \cite{Chenoweth2008} & 1.595 &  $1.1\cdot10^{-6}$ & $-19.17$ & $1.7\cdot10^{-3}$ \\\hline
ReaxFF-C2013 \cite{Srinivasan2015} & 1.356 &  $4.0\cdot10^{-7}$ & $-18.79$ & $5.3\cdot10^{-4}$ \\\hline
\end{tabular}
}
\label{table:bend}
\end{table*}

\begin{table*}
  \centering
    \caption{Fitted parameters $D_{2n}$ of the polynomial approximation  for zigzag CNTs and relative standard errors $d_{2n}$  for REBO-2002 \cite{Brenner2002} and PPBE-G \cite{Wei2011} ($n_\mathrm{max} = 5$ and 4, respectively).}
   \renewcommand{\arraystretch}{1.2}
    \resizebox{0.55\textwidth}{!}{
        \begin{tabular}{*{5}{c}}
\hline
 &  REBO-2002   &  & PPBE-G &  \\\hline
$n$ & $D_{2n}$ (eV$\cdot$\AA$^{2n-2}$) & $d_{2n}$  & $D_{2n}$ (eV$\cdot$\AA$^{2n-2}$) & $d_{2n}$  \\\hline
1 & 1.402 & $2.5\cdot10^{-9}$ &  2.271 & $1.6\cdot10^{-7}$ \\\hline
2 & $-8.912$ & $2.3\cdot10^{-4}$ &  $-42.24$ & $2.4\cdot10^{-3}$ \\\hline
3 & $-1.359\cdot10^{4}$ & $6.9\cdot10^{-2}$ &  $-5.150\cdot10^{5}$ & $3.5\cdot10^{-2}$ \\\hline
4 &  $1.190\cdot10^{10}$  & $2.1\cdot10^{-2}$ &  $1.1058\cdot10^{11}$ & $1.1\cdot10^{-2}$ \\\hline
5 &  $-4.433\cdot10^{16}$  & $5.6\cdot10^{-3}$ &  - & - \\\hline
\end{tabular}
}
\label{table:bend1}
\end{table*}

\begin{table*}
  \centering
    \caption{Literature data on bending rigidity $D_2$ of graphene based on DFT calculations,  calculations using empirical potentials and experimental measurements.}
   \renewcommand{\arraystretch}{1.2}
    \resizebox{0.8\textwidth}{!}{
        \begin{tabular}{*{5}{c}}
\hline
Method & Calculation & System & $D$ (eV)  & Ref. \\\hline
Tersoff & energy-curvature  & CNTs & 1.02 & \cite{Tersoff1992}\\\hline
REBO-1990 & energy-curvature & armchair/zigzag CNTs & 0.85 & \cite{Robertson1992,Yakobson1996}\\\hline
REBO-1990 & \begin{tabular}{@{}c@{}}  analytical limit \\ \& energy-radius  \end{tabular} & \begin{tabular}{@{}c@{}}   \\ armchair/zigzag CNTs \end{tabular} & 0.797 & \cite{Arroyo2004}\\\hline
REBO-1990 & analytical limit & & 0.795 & \cite{Huang2006}\\\hline
REBO-1990 & \begin{tabular}{@{}c@{}}  analytical limit \\ \& energy-radius  \end{tabular} & \begin{tabular}{@{}c@{}}   \\ armchair/zigzag CNTs \end{tabular}  & 0.83 & \cite{Lu2009}\\\hline
AIREBO & MD at 300 K & graphene flakes & 2.83--4.04& \cite{Shen2010}  \\\hline
REBO-2002 &  \begin{tabular}{@{}c@{}}  analytical limit \\ \& energy-radius  \end{tabular} & \begin{tabular}{@{}c@{}}   \\ armchair/zigzag CNTs \end{tabular} & 1.4 & \cite{Lu2009}\\\hline
REBO-2002 & MD at 330 K & ripples in graphene & 0.9 & \cite{Liu2009} \\\hline
REBO-2002 & extrapolation to 0 K & ripples in graphene & 1.12 & \cite{Liu2009} \\\hline
PPBE-G & MD at 300 K & ripples in graphene & 1.56 & \cite{Wei2011} \\\hline
ReaxFF-CHO  & energy-curvature& graphene layer   & 2.1 & \cite{Cranford2011}  \\\hline
ReaxFF-C2013  & energy-curvature & graphene ribbons & 1.31$^\mathrm{a}$ & \cite{Gonzalez2018}  \\\hline
PBE & energy-curvature& armchair CNTs &   1.52 &  \cite{Siahlo2018} \\\hline
PBE & energy-curvature& armchair/zigzag CNTs & 1.5 & \cite{Kudin2001} \\\hline
PBE & energy-curvature& armchair/zigzag CNTs &  1.44--1.52 & \cite{Lebedeva2012b} \\\hline
PBE & energy-curvature& armchair/zigzag/chiral CNTs &  1.44 & \cite{Wei2013} \\\hline
PW91 & energy-curvature& armchair CNTs &  1.52--1.63 & \cite{Cherian2007} \\\hline
PW91 & energy-curvature& zigzag CNTs &  1.66 & \cite{Gulseren2002} \\\hline
LDA & energy-curvature& armchair CNTs &  1.60$^\mathrm{b}$ & \cite{Kurti1998} \\\hline
LDA & energy-curvature&  zigzag CNTs & 1.49$^\mathrm{b}$ &  \cite{Kurti1998} \\\hline
LDA & energy-curvature& armchair CNTs & 1.49 & \cite{Sanchez1999} \\\hline
LDA & energy-curvature& (8,4) CNT & 1.61 & \cite{Sanchez1999} \\\hline
LDA & energy-curvature& (10,0) CNT & 1.61 & \cite{Sanchez1999} \\\hline
LDA & phonon spectrum & graphene & 1.7 & \cite{Sanchez1999} \\\hline
Exp. & phonon spectrum & graphite & 1.2 & \cite{Nicklow1972}\\\hline
\multicolumn{5}{l}{$^\mathrm{a}$Extrapolation to periodic graphene.}  \\
\multicolumn{5}{l}{$^\mathrm{b}$According to the analysis from Ref. \cite{Siahlo2018} and using the bond length of 1.42 \AA~for graphene.}
\end{tabular}
}
\label{table:lit}
\end{table*}

\section{Conclusions}
\label{}
The Tersoff,  REBO-1990, REBO-2000, AIREBO, REBO-2002, LCBOP,  PPBE-G, ReaxFF-CHO and ReaxFF-C2013 were tested with respect to the elastic properties of graphene using the same computational procedure. The elastic properties in the limit of zero strain, such as Young's modulus, Poisson's ratio and bending rigidity, were compared with the experimental and \textit{ab initio} data. The third-order elastic modulus $Y_3$ for uniaxial stretching was also considered. Additional DFT calculations were performed to provide the reference data for the dependences of the effective Young's modulus and Poisson's ratio on the elongation. 

It was revealed that the elastic response of all the potentials considered is non-linear already at elongations of 3\%. This non-linearity, however, is particularly striking for REBO-2000, AIREBO and REBO-2002. It can be responsible for very different results on the Young's modulus and Poisson's ratio evaluated using different intervals of strains and computational approaches. Such a non-linear response and positive third-order elastic modulus are inconsistent with the available experimental and DFT data. A number of artefacts were also discovered for the ReaxFF-C2013 potential, such as a jump discontinuity in the dependence of the Poisson's ratio on the elongation in the armchair direction, non-monotonic dependence of the Poisson's ratio, \textit{etc.} Also the ReaxFF potentials give extremely large Poisson's ratios, while the Tersoff potential predicts a negative value. The best agreement with the reference experimental and \textit{ab initio} data on the in-plane elastic properties at zero elongation is achieved for the LCBOP potential, which captures properly not only the Young's modulus and Poisson's ratio but also the third-order elastic modulus $Y_3$. Correspondingly LCBOP gives qualitatively correct  dependences of the effective Young's modulus and Poisson's ratio on the elongation up to 3\%. Quailitative agreement is also observed for the PPBE-G potential, which is very cheap computationally and is optimal for the use with large systems.

LCBOP, however, fails to reproduce the bending rigidity of graphene in static energy calculations for CNTs. PPBE-G, though realistically describes ripples in graphene at room temperature \cite{Wei2011}, strongly overestimates elastic energies of CNTs. Studies of the temperature dependence of the bending rigidity for this potential are required to clarify the reasons of this discrepancy. The values of the bending rigidity provided by REBO-2000, REBO-2002, ReaxFF-CHO and ReaxFF-C2013 agree the best with  the experimental and \textit{ab initio} data. 

Therefore, none of the considered potential describes adequately both in-plane and out-of-plane deformations of graphene. MD simulations \cite{Fasolino2007} show that the newer version of the LCBOP potential, LCBOPII \cite{Los2005}, with the torsional terms included performs well for ripples in graphene and this might be a good choice for simulations where elastic properties of graphene are important. Nevertheless, further work is still needed on improvement of the accuracy of existing potentials and development of new ones.

\section{Appendix}
The elastic energy of graphene can be also expressed through the elastic constants $C_{ij}$, $C_{ijk}$, $C_{ijkl}$, \textit{etc.}:
\begin{equation} \label{eq_C}
\begin{split}
\Delta U=&\frac{1}{2!} \sum_{i,j=1,2}  C_{ij}\epsilon_i \epsilon_j +\frac{1}{3!} \sum_{i,j,k=1,2}  C_{ijk}\epsilon_i \epsilon_j \epsilon_k \\& +\frac{1}{4!} \sum_{i,j,k,l=1,2}  C_{ijkl}\epsilon_i \epsilon_j \epsilon_k\epsilon_l+...,
\end{split}
\end{equation}
where we choose that index 1 corresponds to the zigzag direction and 2 to the armchair one. Note that due to the rotational symmetry of graphene $C_{11}=C_{22}$. 

The following relations hold between the Poisson's ratio $\nu_0$ in the limit of zero elongation, Young's modulus $Y_2$ and elastic constants \cite{Cao2014}:
\begin{equation} \label{eq_C1}
\nu_0 = C_{12}/C_{11},
\end{equation}
\begin{equation} \label{eq_C2}
Y_2 = C_{11} \left(1-\nu_0^2\right).
\end{equation}
The third-order modulus $Y_3$ for the zigzag direction is given by (see Ref. \cite{Cao2014}):
\begin{equation} \label{eq_C3}
Y_3 = C_{111}-3C_{112}\nu_0+3C_{122}\nu_0^2-C_{222}\nu_0^3.
\end{equation}
The expression for the armchair direction is obtained by interchanging the indices 1 and 2. Note that for graphene $C_{122}=C_{111}-C_{222}+C_{112}$ \cite{Cao2014,Wei2009}.

\section*{Acknowledgments}
We thank Prof. Feng Wang for providing the LAMMPS script for the PPBE-G potential. AMP and AAK acknowledge the Russian Foundation for Basic Research (Grant 18-02-00985). IVL acknowledges Grupos Consolidados del Gobierno Vasco (IT-578-13). This work has been carried out using computing resources of the federal collective usage center Complex for Simulation and Data Processing for Mega-science Facilities at NRC ``Kurchatov Institute", http://ckp.nrcki.ru/.

\section*{Data availability}
The raw data required to reproduce these findings are available to download from https://data.mendeley.com/datasets/wfpn7t3bxh/1 (data obtained using interatomic potentials) and https://data.mendeley.com/datasets/x2pd23dv4n/1 (DFT data). The LAMMPS scripts, input and output files for the comparison of computational performance of the potentials are available to download from https://data.mendeley.com/datasets/nnd9d4y2mb/2.

\section*{References}
\bibliography{cms}

\begin{thebibliography}{10}
\expandafter\ifx\csname url\endcsname\relax
  \def\url#1{\texttt{#1}}\fi
\expandafter\ifx\csname urlprefix\endcsname\relax\def\urlprefix{URL }\fi
\expandafter\ifx\csname href\endcsname\relax
  \def\href#1#2{#2} \def\path#1{#1}\fi

\bibitem{Tersoff1988}
J.~Tersoff, Empirical interatomic potential for carbon, with applications to
  amorphous carbon, Phys. Rev. Lett. 61 (1988) 2879--2882.
\newblock \href {http://dx.doi.org/10.1103/PhysRevLett.61.2879}
  {\path{doi:10.1103/PhysRevLett.61.2879}}.

\bibitem{Brenner1990}
D.~W. Brenner, Empirical potential for hydrocarbons for use in simulating the
  chemical vapor deposition of diamond films, Phys. Rev. B 42 (1990)
  9458--9471.
\newblock \href {http://dx.doi.org/10.1103/PhysRevB.42.9458}
  {\path{doi:10.1103/PhysRevB.42.9458}}.

\bibitem{Stuart2000}
S.~J. Stuart, A.~B. Tutein, J.~A. Harrison, A reactive potential for
  hydrocarbons with intermolecular interactions, J. Chem. Phys. 112~(14) (2000)
  6472--6486.
\newblock \href {http://dx.doi.org/10.1063/1.481208}
  {\path{doi:10.1063/1.481208}}.

\bibitem{Brenner2002}
D.~W. Brenner, O.~A. Shenderova, J.~A. Harrison, S.~J. Stuart, B.~Ni, S.~B.
  Sinnott, A second-generation reactive empirical bond order ({REBO}) potential
  energy expression for hydrocarbons, J. Phys.: Condens. Matter 14 (2002)
  783--802.
\newblock \href {http://dx.doi.org/10.1088/0953-8984/14/4/312}
  {\path{doi:10.1088/0953-8984/14/4/312}}.

\bibitem{Los2003}
J.~H. Los, A.~Fasolino, Intrinsic long-range bond-order potential for carbon:
  Performance in {M}onte {C}arlo simulations of graphitization, Phys. Rev. B 68
  (2003) 024107.
\newblock \href {http://dx.doi.org/10.1103/PhysRevB.68.024107}
  {\path{doi:10.1103/PhysRevB.68.024107}}.

\bibitem{Wei2011}
D.~Wei, Y.~Song, F.~Wang, A simple molecular mechanics potential for $\mu$m
  scale graphene simulations from the adaptive force matching method, J. Chem.
  Phys. 134 (2011) 184704.
\newblock \href {http://dx.doi.org/10.1063/1.3589163}
  {\path{doi:10.1063/1.3589163}}.

\bibitem{Chenoweth2008}
K.~Chenoweth, A.~C.~T. van Duin, W.~A. Goddard, {ReaxFF} reactive force field
  for molecular dynamics simulations of hydrocarbon oxidation, J. Phys. Chem. A
  112 (2008) 1040--1053.
\newblock \href {http://dx.doi.org/10.1021/jp709896w}
  {\path{doi:10.1021/jp709896w}}.

\bibitem{Srinivasan2015}
S.~G. Srinivasan, A.~C.~T. van Duin, P.~Ganesh, Development of a {ReaxFF}
  potential for carbon condensed phases and its application to the thermal
  fragmentation of a large fullerene, J. Phys. Chem. A 119 (2015) 571--580.
\newblock \href {http://dx.doi.org/10.1021/jp510274e}
  {\path{doi:10.1021/jp510274e}}.

\bibitem{Zhu2011}
J.~Zhu, D.~Shi, The integrated effects of temperature and stress on the
  formation of carbon linear atomic chains from graphene nanoribbons, J. Appl.
  Phys. 110 (2011) 104311.
\newblock \href {http://dx.doi.org/10.1063/1.3662183}
  {\path{doi:10.1063/1.3662183}}.

\bibitem{Sun2013}
Y.~J. Sun, F.~Ma, Y.~H. Huang, T.~W. Hu, K.~W. Xu, P.~K. Chu, Effects of
  loading mode and orientation on deformation mechanism of graphene
  nanoribbons, Appl. Phys. Lett. 103 (2013) 191906.
\newblock \href {http://dx.doi.org/10.1063/1.4829480}
  {\path{doi:10.1063/1.4829480}}.

\bibitem{Lebedeva2008}
I.~V. Lebedeva, A.~A. Knizhnik, A.~A.Bagatur{'}yants, B.~V. Potapkin, Kinetics
  of {2D--3D} transformations of carbon nanostructures, Physica E 40 (2008)
  2589--2595.
\newblock \href {http://dx.doi.org/0.1016/j.physe.2007.09.155}
  {\path{doi:0.1016/j.physe.2007.09.155}}.

\bibitem{Lebedeva2012}
I.~V. Lebedeva, A.~A. Knizhnik, A.~M. Popov, B.~V. Potapkin, Ni-assisted
  transformation of graphene flakes to fullerenes, J. Phys. Chem. C 116 (2012)
  6572--6584.
\newblock \href {http://dx.doi.org/10.1021/jp212165g}
  {\path{doi:10.1021/jp212165g}}.

\bibitem{Santana2013}
A.~Santana, A.~Zobelli, J.~Kotakoski, A.~Chuvilin, E.~Bichoutskaia, Inclusion
  of radiation damage dynamics in high-resolution transmission electron
  microscopy image simulations: The example of graphene, Phys. Rev. B 87 (2013)
  094110.
\newblock \href {http://dx.doi.org/10.1103/PhysRevB.87.094110}
  {\path{doi:10.1103/PhysRevB.87.094110}}.

\bibitem{Skowron2013}
S.~T. Skowron, I.~V. Lebedeva, A.~M. Popov, E.~Bichoutskaia, Approaches to
  modelling irradiation-induced processes in transmission electron microscopy,
  Nanoscale 5 (2013) 6677--6692.
\newblock \href {http://dx.doi.org/10.1039/C3NR02130K}
  {\path{doi:10.1039/C3NR02130K}}.

\bibitem{Yamaletdinov2017}
R.~D. Yamaletdinov, Y.~V. Pershin, Finding stable graphene conformations from
  pull and release experiments with molecular dynamics, Sci. Rep. 7 (2017)
  42356.
\newblock \href {http://dx.doi.org/10.1038/srep42356}
  {\path{doi:10.1038/srep42356}}.

\bibitem{Shi2010}
X.~Shi, N.~M. Pugno, H.~Gao, Tunable core size of carbon nanoscrolls, J.
  Comput. Theor. Nanosci. 7 (2010) 517--521.
\newblock \href {http://dx.doi.org/10.1166/jctn.2010.1387}
  {\path{doi:10.1166/jctn.2010.1387}}.

\bibitem{Wang2015}
Y.~Wang, H.~F. Zhan, C.~Yang, Y.~Xiang, Y.~Y. Zhang, Formation of carbon
  nanoscrolls from graphene nanoribbons: A molecular dynamics study, Comp. Mat.
  Sci. 96 (2015) 300--305.
\newblock \href {http://dx.doi.org/10.1016/j.commatsci.2014.09.039}
  {\path{doi:10.1016/j.commatsci.2014.09.039}}.

\bibitem{Shen2012}
H.~J. Shen, K.~Cheng, Tensile properties and thermal conductivity of graphene
  nanoribbons encapsulated in single-walled carbon nanotube, Molecular
  Simulation 38 (2012) 922--927.
\newblock \href {http://dx.doi.org/10.1080/08927022.2012.672739}
  {\path{doi:10.1080/08927022.2012.672739}}.

\bibitem{Furuhashi2013}
F.~Furuhashi, K.~Shintani, Morphology of a graphene nanoribbon encapsulated in
  a carbon nanotube, AIP Advances 3 (2013) 092103.
\newblock \href {http://dx.doi.org/10.1063/1.4821102}
  {\path{doi:10.1063/1.4821102}}.

\bibitem{Li2015}
Y.~Li, W.~Chen, H.~Ren, X.~Zhou, H.~Li, Multiple helical configuration and
  quantity threshold of graphene nanoribbons inside a single-walled carbon
  nanotube, Sci. Rep. 5 (2015) 13741.
\newblock \href {http://dx.doi.org/10.1038/srep13741}
  {\path{doi:10.1038/srep13741}}.

\bibitem{Fang2015}
T.~H. Fang, W.-J. Chang, Y.-L. Feng, Mechanical characteristics of graphene
  nanoribbons encapsulated in single-walled carbon nanotubes using molecular
  dynamics simulations, Appl. Surf. Sci. 356 (2015) 221--225.
\newblock \href {http://dx.doi.org/10.1016/j.apsusc.2015.07.210}
  {\path{doi:10.1016/j.apsusc.2015.07.210}}.

\bibitem{Popov2011}
A.~M. Popov, I.~V. Lebedeva, A.~A. Knizhnik, Y.~E. Lozovik, B.~V. Potapkin,
  Commensurate-incommensurate phase transition in bilayer graphene, Phys. Rev.
  B 84 (2011) 045404.
\newblock \href {http://dx.doi.org/10.1103/PhysRevB.84.045404}
  {\path{doi:10.1103/PhysRevB.84.045404}}.

\bibitem{Argentero2017}
G.~Argentero, A.~Mittelberger, M.~R.~A. M., Y.~Cao, T.~J. Pennycook,
  C.~Mangler, C.~Kramberger, J.~Kotakoski, A.~K. Geim, J.~C. Meyer, Unraveling
  the {3D} atomic structure of a suspended graphene/{hBN} van der {W}aals
  heterostructure, Nano Lett. 17 (2017) 1409–1416.
\newblock \href {http://dx.doi.org/10.1021/acs.nanolett.6b04360}
  {\path{doi:10.1021/acs.nanolett.6b04360}}.

\bibitem{Wang2009}
C.~Y. Wang, K.~Mylvaganam, L.~C. Zhang, Wrinkling of monolayer graphene: A
  study by molecular dynamics and continuum plate theory, Phys. Rev. B 80
  (2009) 155445.
\newblock \href {http://dx.doi.org/10.1103/PhysRevB.80.155445}
  {\path{doi:10.1103/PhysRevB.80.155445}}.

\bibitem{Popov2011x}
A.~M. Popov, I.~V. Lebedeva, A.~A. Knizhnik, Y.~E. Lozovik, B.~V. Potapkin,
  Molecular dynamics simulation of the self-retracting motion of a graphene
  flake, Phys. Rev. B 84 (2011) 245437.
\newblock \href {http://dx.doi.org/10.1103/PhysRevB.84.245437}
  {\path{doi:10.1103/PhysRevB.84.245437}}.

\bibitem{Lebedeva2012a}
I.~V. Lebedeva, A.~A. Knizhnik, A.~M. Popov, Y.~E. Lozovik, B.~V. Potapkin,
  Modeling of graphene-based {NEMS}, Physica E 44 (2012) 949--954.
\newblock \href {http://dx.doi.org/10.1016/j.physe.2011.07.018}
  {\path{doi:10.1016/j.physe.2011.07.018}}.

\bibitem{Lebedeva2011}
I.~V. Lebedeva, A.~A. Knizhnik, A.~M. Popov, Y.~E. Lozovik, B.~V. Potapkin,
  Interlayer interaction and relative vibrations of bilayer graphene, Phys.
  Chem. Chem. Phys. 13 (2011) 5687--5695.
\newblock \href {http://dx.doi.org/10.1039/c0cp02614j}
  {\path{doi:10.1039/c0cp02614j}}.

\bibitem{Shi2010a}
X.~Shi, Y.~Cheng, N.~M. Pugno, H.~Gao, A translational nanoactuator based on
  carbon nanoscrolls on substrates, Appl. Phys. Lett. 96 (2010) 053115.
\newblock \href {http://dx.doi.org/10.1063/1.3302284}
  {\path{doi:10.1063/1.3302284}}.

\bibitem{Hwang2013}
Z.~Hwang, J.~Lee, J.~W. Kang, Molecular dynamics study on graphene-based
  nanoelectromechanical relays, J. Comput. Theor. Nanosci. 10 (2013)
  1892--1898.
\newblock \href {http://dx.doi.org/10.1166/jctn.2013.31}
  {\path{doi:10.1166/jctn.2013.31}}.

\bibitem{Hwang2014}
H.~J. Hwang, J.~W. Kang, Nonvolatile graphene nanoflake shuttle memory, Physica
  E 56 (2014) 17--23.
\newblock \href {http://dx.doi.org/10.1016/j.physe.2013.08.009}
  {\path{doi:10.1016/j.physe.2013.08.009}}.

\bibitem{Kang2014}
J.~W. Kang, J.~Park, O.~K. Kwon, Developing a nanoelectromechanical shuttle
  graphene-nanoflake device, Physica E 58 (2014) 88--93.
\newblock \href {http://dx.doi.org/10.1016/j.physe.2013.12.001}
  {\path{doi:10.1016/j.physe.2013.12.001}}.

\bibitem{Kang2016}
J.~W. Kang, R.~Oh, K.-S. Kim, O.-K. Kwon, Molecular dynamics study on
  nanoelectromechanical graphene nanoribbon device with graphene nanoflake
  shuttle, J. Nanosci. Nanotechnol. 16 (2016) 11975–--11979.
\newblock \href {http://dx.doi.org/10.1166/jnn.2016.13628}
  {\path{doi:10.1166/jnn.2016.13628}}.

\bibitem{Berinskii2010}
I.~E. Berinskii, A.~M. Krivtsov, On using many-particle interatomic potentials
  to compute elastic properties of graphene and diamond, Mechanics of Solids 45
  (2010) 815--834.
\newblock \href {http://dx.doi.org/10.3103/S0025654410060063}
  {\path{doi:10.3103/S0025654410060063}}.

\bibitem{Sgouros2016}
A.~P. Sgouros, G.~Kalosakas, C.~Galiotis, K.~Papagelis, Uniaxial compression of
  suspended single and multilayer graphenes, 2D Materials 3~(2) (2016) 025033.

\bibitem{Memarian2015}
F.~Memarian, A.~Fereidoon, M.~D. Ganji, Graphene {Y}oung's modulus: Molecular
  mechanics and {DFT} treatments, Superlattices and Microstructures 85 (2015)
  348--356.
\newblock \href {http://dx.doi.org/10.1016/j.spmi.2015.06.001}
  {\path{doi:10.1016/j.spmi.2015.06.001}}.

\bibitem{Arroyo2004}
M.~Arroyo, T.~Belytschko, Finite crystal elasticity of carbon nanotubes based
  on the exponential cauchy-born rule, Phys. Rev. B 69 (2004) 115415.
\newblock \href {http://dx.doi.org/10.1103/PhysRevB.69.115415}
  {\path{doi:10.1103/PhysRevB.69.115415}}.

\bibitem{Huang2006}
Y.~Huang, J.~Wu, K.~C. Hwang, Thickness of graphene and single-wall carbon
  nanotubes, Phys. Rev. B 74 (2006) 245413.
\newblock \href {http://dx.doi.org/10.1103/PhysRevB.74.245413}
  {\path{doi:10.1103/PhysRevB.74.245413}}.

\bibitem{Reddy2006}
C.~D. Reddy, S.~Rajendran, K.~M. Liew, Equilibrium configuration and continuum
  elastic properties of finite sized graphene, Nanotechnology 17 (2006) 864.

\bibitem{Shen2010}
L.~Shen, H.-S. Shen, C.-L. Zhang, Temperature-dependent elastic properties of
  single layer graphene sheets, Materials \& Design 31 (2010) 4445--4449.
\newblock \href {http://dx.doi.org/10.1016/j.matdes.2010.04.016}
  {\path{doi:10.1016/j.matdes.2010.04.016}}.

\bibitem{Zhao2009}
H.~Zhao, K.~Min, N.~R. Aluru, Size and chirality dependent elastic properties
  of graphene nanoribbons under uniaxial tension, Nano Lett. 9 (2009)
  3012--3015.
\newblock \href {http://dx.doi.org/10.1021/nl901448z}
  {\path{doi:10.1021/nl901448z}}.

\bibitem{WenXing2004}
B.~WenXing, Z.~ChangChun, C.~WanZhao, Simulation of {Y}oung's modulus of
  single-walled carbon nanotubes by molecular dynamics, Phys. B: Condens. Mat.
  352 (2004) 156--163.
\newblock \href {http://dx.doi.org/10.1016/j.physb.2004.07.005}
  {\path{doi:10.1016/j.physb.2004.07.005}}.

\bibitem{Jensen2015}
B.~D. Jensen, K.~E. Wise, G.~M. Odegard, Simulation of the elastic and ultimate
  tensile properties of diamond, graphene, carbon nanotubes, and amorphous
  carbon using a revised {ReaxFF} parametrization, J. Phys. Chem. A 119 (2015)
  9710--9721.
\newblock \href {http://dx.doi.org/10.1021/acs.jpca.5b05889}
  {\path{doi:10.1021/acs.jpca.5b05889}}.

\bibitem{Cranford2011}
S.~Cranford, M.~J. Buehler, Twisted and coiled ultralong multilayer graphene
  ribbons, Modelling and Simulation in Materials Science and Engineering 19
  (2011) 054003.

\bibitem{Tersoff1992}
J.~Tersoff, Energies of fullerenes, Phys. Rev. B 46 (1992) 15546--15549.
\newblock \href {http://dx.doi.org/10.1103/PhysRevB.46.15546}
  {\path{doi:10.1103/PhysRevB.46.15546}}.

\bibitem{Cao2014}
G.~Cao, Atomistic studies of mechanical properties of graphene, Polymers 6
  (2014) 2404--2432.
\newblock \href {http://dx.doi.org/10.3390/polym6092404}
  {\path{doi:10.3390/polym6092404}}.

\bibitem{Robertson1992}
D.~H. Robertson, D.~W. Brenner, J.~W. Mintmire, Energetics of nanoscale
  graphitic tubules, Phys. Rev. B 45 (1992) 12592--12595.
\newblock \href {http://dx.doi.org/10.1103/PhysRevB.45.12592}
  {\path{doi:10.1103/PhysRevB.45.12592}}.

\bibitem{Yakobson1996}
B.~I. Yakobson, C.~J. Brabec, J.~Bernholc, Nanomechanics of carbon tubes:
  Instabilities beyond linear response, Phys. Rev. Lett. 76 (1996) 2511--2514.
\newblock \href {http://dx.doi.org/10.1103/PhysRevLett.76.2511}
  {\path{doi:10.1103/PhysRevLett.76.2511}}.

\bibitem{Lu2009}
Q.~Lu, M.~Arroyo, R.~Huang, Elastic bending modulus of monolayer graphene, J.
  Phys. D: Appl. Phys. 42 (2009) 102002.

\bibitem{Gonzalez2018}
R.~I. Gonz\'{a}lez, F.~J. Valencia, J.~Rogan, J.~A. Valdivia, J.~Sofo, M.~Kiwi,
  F.~Munoz, Bending energy of {2D} materials: graphene{,} {MoS}$_2$ and
  imogolite, RSC Adv. 8 (2018) 4577--4583.
\newblock \href {http://dx.doi.org/10.1039/C7RA10983K}
  {\path{doi:10.1039/C7RA10983K}}.

\bibitem{Liu2009}
P.~Liu, Y.~W. Zhang, Temperature-dependent bending rigidity of graphene, Appl.
  Phys. Lett. 94 (2009) 231912.
\newblock \href {http://dx.doi.org/10.1063/1.3155197}
  {\path{doi:10.1063/1.3155197}}.

\bibitem{Lee2008}
C.~Lee, X.~Wei, J.~W. Kysar, J.~Hone, Measurement of the elastic properties and
  intrinsic strength of monolayer graphene, Science 321 (2008) 385--388.
\newblock \href {http://dx.doi.org/10.1126/science.1157996}
  {\path{doi:10.1126/science.1157996}}.

\bibitem{Bosak2007}
A.~Bosak, M.~Krisch, M.~Mohr, J.~Maultzsch, C.~Thomsen, Elasticity of
  single-crystalline graphite: Inelastic x-ray scattering study, Phys. Rev. B
  75 (2007) 153408.
\newblock \href {http://dx.doi.org/10.1103/PhysRevB.75.153408}
  {\path{doi:10.1103/PhysRevB.75.153408}}.

\bibitem{Blakslee1970}
O.~L. Blakslee, D.~G. Proctor, E.~J. Seldin, G.~B. Spence, T.~Weng, Elastic
  constants of compression-annealed pyrolytic graphite, J. Appl. Phys. 41
  (1970) 3373--3382.
\newblock \href {http://dx.doi.org/10.1063/1.1659428}
  {\path{doi:10.1063/1.1659428}}.

\bibitem{Andres2012}
P.~L. de~Andres, F.~Guinea, M.~I. Katsnelson, Density functional theory
  analysis of flexural modes, elastic constants, and corrugations in strained
  graphene, Phys. Rev. B 86 (2012) 245409.
\newblock \href {http://dx.doi.org/10.1103/PhysRevB.86.245409}
  {\path{doi:10.1103/PhysRevB.86.245409}}.

\bibitem{Nicklow1972}
R.~Nicklow, N.~Wakabayashi, H.~G. Smith, Lattice dynamics of pyrolytic
  graphite, Phys. Rev. B 5 (1972) 4951--4962.
\newblock \href {http://dx.doi.org/10.1103/PhysRevB.5.4951}
  {\path{doi:10.1103/PhysRevB.5.4951}}.

\bibitem{Wei2009}
X.~Wei, B.~Fragneaud, C.~A. Marianetti, J.~W. Kysar, Nonlinear elastic behavior
  of graphene: \emph{Ab initio} calculations to continuum description, Phys.
  Rev. B 80 (2009) 205407.
\newblock \href {http://dx.doi.org/10.1103/PhysRevB.80.205407}
  {\path{doi:10.1103/PhysRevB.80.205407}}.

\bibitem{Kalosakas2013}
G.~Kalosakas, N.~N. Lathiotakis, C.~Galiotis, K.~Papagelis, In-plane force
  fields and elastic properties of graphene, J. Appl. Phys. 113 (2013) 134307.
\newblock \href {http://dx.doi.org/10.1063/1.4798384}
  {\path{doi:10.1063/1.4798384}}.

\bibitem{Cadelano2012}
E.~Cadelano, L.~Colombo, Effect of hydrogen coverage on the {Y}oung's modulus
  of graphene, Phys. Rev. B 85 (2012) 245434.
\newblock \href {http://dx.doi.org/10.1103/PhysRevB.85.245434}
  {\path{doi:10.1103/PhysRevB.85.245434}}.

\bibitem{Kudin2001}
K.~N. Kudin, G.~E. Scuseria, B.~I. Yakobson, {C$_2$F}, {BN}, and {C} nanoshell
  elasticity from \emph{ab initio} computations, Phys. Rev. B 64 (2001) 235406.
\newblock \href {http://dx.doi.org/10.1103/PhysRevB.64.235406}
  {\path{doi:10.1103/PhysRevB.64.235406}}.

\bibitem{Mounet2005}
N.~Mounet, N.~Marzari, First-principles determination of the structural,
  vibrational and thermodynamic properties of diamond, graphite, and
  derivatives, Phys. Rev. B 71 (2005) 205214.
\newblock \href {http://dx.doi.org/10.1103/PhysRevB.71.205214}
  {\path{doi:10.1103/PhysRevB.71.205214}}.

\bibitem{Lebedeva2016}
I.~V. Lebedeva, A.~V. Lebedev, A.~M. Popov, A.~A. Knizhnik, Dislocations in
  stacking and commensurate-incommensurate phase transition in bilayer graphene
  and hexagonal boron nitride, Phys. Rev. B 93 (2016) 235414.
\newblock \href {http://dx.doi.org/10.1103/PhysRevB.93.235414}
  {\path{doi:10.1103/PhysRevB.93.235414}}.

\bibitem{Liu2007}
F.~Liu, P.~Ming, J.~Li, Ab initio calculation of ideal strength and phonon
  instability of graphene under tension, Phys. Rev. B 76 (2007) 064120.
\newblock \href {http://dx.doi.org/10.1103/PhysRevB.76.064120}
  {\path{doi:10.1103/PhysRevB.76.064120}}.

\bibitem{Zhou2016}
L.~Zhou, G.~Cao, Nonlinear anisotropic deformation behavior of a graphene
  monolayer under uniaxial tension, Phys. Chem. Chem. Phys. 18 (2016)
  1657--1664.
\newblock \href {http://dx.doi.org/10.1039/C5CP05791D}
  {\path{doi:10.1039/C5CP05791D}}.

\bibitem{Gui2008}
G.~Gui, J.~Li, J.~Zhong, Band structure engineering of graphene by strain:
  First-principles calculations, Phys. Rev. B 78 (2008) 075435.
\newblock \href {http://dx.doi.org/10.1103/PhysRevB.78.075435}
  {\path{doi:10.1103/PhysRevB.78.075435}}.

\bibitem{Shao2012}
T.~Shao, B.~Wen, R.~Melnik, S.~Yao, Y.~Kawazoe, Y.~Tian, Temperature dependent
  elastic constants and ultimate strength of graphene and graphyne, J. Chem.
  Phys. 137 (2012) 194901.
\newblock \href {http://dx.doi.org/10.1063/1.4766203}
  {\path{doi:10.1063/1.4766203}}.

\bibitem{Savini2011}
G.~Savini, Y.~J. Dappe, S.~\, Bending modes, elastic constants and mechanical
  stability of graphitic systems, Carbon 49 (2011) 62--69.
\newblock \href {http://dx.doi.org/10.1016/j.carbon.2010.08.042}
  {\path{doi:10.1016/j.carbon.2010.08.042}}.

\bibitem{Siahlo2018}
A.~I. Siahlo, N.~A. Poklonski, A.~V. Lebedev, I.~V. Lebedeva, A.~M. Popov,
  S.~A. Vyrko, A.~A. Knizhnik, Y.~E. Lozovik, Structure and energetics of
  carbon, hexagonal boron nitride, and carbon/hexagonal boron nitride
  single-layer and bilayer nanoscrolls, Phys. Rev. Materials 2 (2018) 036001.
\newblock \href {http://dx.doi.org/10.1103/PhysRevMaterials.2.036001}
  {\path{doi:10.1103/PhysRevMaterials.2.036001}}.

\bibitem{Lebedeva2012b}
I.~V. Lebedeva, A.~M. Popov, A.~A. Knizhnik, A.~N. Khlobystov, B.~V. Potapkin,
  Chiral graphene nanoribbon inside a carbon nanotube: \emph{ab initio} study,
  Nanoscale 4 (2012) 4522--4529.
\newblock \href {http://dx.doi.org/10.1039/C2NR30144J}
  {\path{doi:10.1039/C2NR30144J}}.

\bibitem{Wei2013}
Y.~Wei, B.~Wang, J.~Wu, R.~Yang, M.~L. Dunn, Bending rigidity and gaussian
  bending stiffness of single-layered graphene, Nano Letters 13 (2013) 26--30.
\newblock \href {http://dx.doi.org/10.1021/nl303168w}
  {\path{doi:10.1021/nl303168w}}.

\bibitem{Cherian2007}
R.~Cherian, P.~Mahadevan, Elastic properties of carbon nanotubes: An atomistic
  approach, J. Nanosci. Nanotechnol. 7 (2007) 1779--1782.
\newblock \href {http://dx.doi.org/10.1166/jnn.2007.714}
  {\path{doi:10.1166/jnn.2007.714}}.

\bibitem{Gulseren2002}
O.~G\"ulseren, T.~Yildirim, S.~Ciraci, Systematic \emph{ab initio} study of
  curvature effects in carbon nanotubes, Phys. Rev. B 65 (2002) 153405.
\newblock \href {http://dx.doi.org/10.1103/PhysRevB.65.153405}
  {\path{doi:10.1103/PhysRevB.65.153405}}.

\bibitem{Kurti1998}
J.~K\"urti, G.~Kresse, H.~Kuzmany, First-principles calculations of the radial
  breathing mode of single-wall carbon nanotubes, Phys. Rev. B 58 (1998)
  R8869--R8872.
\newblock \href {http://dx.doi.org/10.1103/PhysRevB.58.R8869}
  {\path{doi:10.1103/PhysRevB.58.R8869}}.

\bibitem{Sanchez1999}
D.~S\'anchez-Portal, E.~Artacho, J.~M. Soler, A.~Rubio, P.~Ordej\'on, \emph{Ab
  initio} structural, elastic, and vibrational properties of carbon nanotubes,
  Phys. Rev. B 59 (1999) 12678--12688.
\newblock \href {http://dx.doi.org/10.1103/PhysRevB.59.12678}
  {\path{doi:10.1103/PhysRevB.59.12678}}.

\bibitem{Ceperley1980}
D.~M. Ceperley, B.~J. Alder, Ground state of the electron gas by a stochastic
  method, Phys. Rev. Lett. 45 (1980) 566--569.
\newblock \href {http://dx.doi.org/10.1103/PhysRevLett.45.566}
  {\path{doi:10.1103/PhysRevLett.45.566}}.

\bibitem{Perdew1996}
J.~P. Perdew, K.~Burke, M.~Ernzerhof, Generalized gradient approximation made
  simple, Phys. Rev. Lett. 77 (1996) 3865--3868.
\newblock \href {http://dx.doi.org/10.1103/PhysRevLett.77.3865}
  {\path{doi:10.1103/PhysRevLett.77.3865}}.

\bibitem{Perdew92}
J.~P. Perdew, Y.~Wang, Pair-distribution function and its coupling-constant
  average for the spin-polarized electron gas, Phys. Rev. B 46 (1992)
  12947--12954.
\newblock \href {http://dx.doi.org/10.1103/PhysRevB.46.12947}
  {\path{doi:10.1103/PhysRevB.46.12947}}.

\bibitem{Grimme2006}
S.~Grimme, Semiempirical {GGA}-type density functional constructed with a
  long-range dispersion correction, J. Comp. Chem. 27 (2006) 1787--1799.
\newblock \href {http://dx.doi.org/10.1002/jcc.20495}
  {\path{doi:10.1002/jcc.20495}}.

\bibitem{Lee2010}
K.~Lee, E.~D. Murray, L.~Kong, B.~I. Lundqvist, D.~C. Langreth, Higher-accuracy
  van der {W}aals density functional, Phys. Rev. B 82 (2010) 081101.
\newblock \href {http://dx.doi.org/10.1103/PhysRevB.82.081101}
  {\path{doi:10.1103/PhysRevB.82.081101}}.

\bibitem{Baroni2001}
S.~Baroni, S.~de~Gironcoli, A.~Dal~Corso, P.~Giannozzi, Phonons and related
  crystal properties from density-functional perturbation theory, Rev. Mod.
  Phys. 73 (2001) 515--562.
\newblock \href {http://dx.doi.org/10.1103/RevModPhys.73.515}
  {\path{doi:10.1103/RevModPhys.73.515}}.

\bibitem{Sinitsa2014}
A.~S. Sinitsa, I.~V. Lebedeva, A.~A. Knizhnik, A.~M. Popov, S.~T. Skowron,
  E.~Bichoutskaia, Formation of nickel-carbon heterofullerenes under electron
  irradiation, Dalton Trans. 43 (2014) 7499--7513.
\newblock \href {http://dx.doi.org/10.1039/C3DT53385A}
  {\path{doi:10.1039/C3DT53385A}}.

\bibitem{Sinitsa2018}
A.~S. Sinitsa, I.~V. Lebedeva, A.~M. Popov, A.~A. Knizhnik, Long triple carbon
  chains formation by heat treatment of graphene nanoribbon: Molecular dynamics
  study with revised {B}renner potential, Carbon 140 (2018) 543--556.
\newblock \href {http://dx.doi.org/10.1016/j.carbon.2018.08.022}
  {\path{doi:10.1016/j.carbon.2018.08.022}}.

\bibitem{Omololu2008}
O.~Akin-Ojo, Y.~Song, F.~Wang, Developing \emph{ab initio} quality force fields
  from condensed phase quantum-mechanics/molecular-mechanics calculations
  through the adaptive force matching method, J. Chem. Phys. 129~(6) (2008)
  064108.
\newblock \href {http://dx.doi.org/10.1063/1.2965882}
  {\path{doi:10.1063/1.2965882}}.

\bibitem{Omololu2011}
A.-O. Omololu, W.~Feng, The quest for the best nonpolarizable water model from
  the adaptive force matching method, J. Comp. Chem. 32 (2011) 453--462.
\newblock \href {http://dx.doi.org/10.1002/jcc.21634}
  {\path{doi:10.1002/jcc.21634}}.

\bibitem{vanDuin2001}
A.~C.~T. van Duin, S.~Dasgupta, F.~Lorant, W.~A. Goddard, {ReaxFF}:  a
  reactive force field for hydrocarbons, J. Phys. Chem. A 105~(41) (2001)
  9396--9409.
\newblock \href {http://dx.doi.org/10.1021/jp004368u}
  {\path{doi:10.1021/jp004368u}}.

\bibitem{MDKMC}
A.~A. Knizhnik, MD-kMC code. Kintech Lab 2000-2018.

\bibitem{Plimpton1995}
S.~Plimpton, Fast parallel algorithms for short-range molecular dynamics, J.
  Comp. Phys. 117 (1995) 1--19.
\newblock \href {http://dx.doi.org/10.1006/jcph.1995.1039}
  {\path{doi:10.1006/jcph.1995.1039}}.

\bibitem{Trucano1975}
P.~Trucano, R.~Chen, Structure of graphite by neutron diffraction, Nature 258
  (1975) 136--137.
\newblock \href {http://dx.doi.org/10.1038/258136a0}
  {\path{doi:10.1038/258136a0}}.

\bibitem{Zhao1989}
Y.~X. Zhao, I.~L. Spain, X-ray diffraction data for graphite to 20 {GPa}, Phys.
  Rev. B 40 (1989) 993--997.
\newblock \href {http://dx.doi.org/10.1103/PhysRevB.40.993}
  {\path{doi:10.1103/PhysRevB.40.993}}.

\bibitem{Ludsteck1972}
V.~A. Ludsteck, Bestimmung der {\aa}nderung der gitterkonstanten und des
  anisotropen {D}ebye-{W}aller-faktors von graphit mittels neutronenbeugung im
  temperaturbereich von 25 bis 1850$^{\circ}$ {C}, Acta Crystallographica,
  Section A 28 (1972) 59--65.
\newblock \href {http://dx.doi.org/10.1107/S0567739472000130}
  {\path{doi:10.1107/S0567739472000130}}.

\bibitem{Baskin1955}
V.~Baskin, L.~Meyer, Lattice constants of graphite at low temperatures, Phys.
  Rev. 100 (1955) 544.
\newblock \href {http://dx.doi.org/10.1103/PhysRev.100.544}
  {\path{doi:10.1103/PhysRev.100.544}}.

\bibitem{Levenberg}
K.~Levenberg, A method for the solution of certain non-linear problems in least
  squares, Quart. Appl. Math. 2 (1944) 164--168.
\newblock \href {http://dx.doi.org/10.1090/qam/10666}
  {\path{doi:10.1090/qam/10666}}.

\bibitem{Marquardt}
D.~Marquardt, An algorithm for least-squares estimation of nonlinear
  parameters, Journal of the Society for Industrial and Applied Mathematics 11
  (1963) 431--441.
\newblock \href {http://dx.doi.org/10.1137/0111030}
  {\path{doi:10.1137/0111030}}.

\bibitem{Kresse1996}
G.~Kresse, J.~Furthm\"{u}ller, Efficient iterative schemes for \textit{ab
  initio} total-energy calculations using a plane-wave basis set, Phys. Rev. B
  54 (1996) 11169--11186.
\newblock \href {http://dx.doi.org/10.1103/PhysRevB.54.11169}
  {\path{doi:10.1103/PhysRevB.54.11169}}.

\bibitem{Kresse1999}
G.~Kresse, D.~Joubert, From ultrasoft pseudopotentials to the projector
  augmented-wave method, Phys. Rev. B 59 (1999) 1758--1775.
\newblock \href {http://dx.doi.org/10.1103/PhysRevB.59.1758}
  {\path{doi:10.1103/PhysRevB.59.1758}}.

\bibitem{Monkhorst1976}
H.~J. Monkhorst, J.~D. Pack, Special points for {B}rillouin-zone integrations,
  Phys. Rev. B 13 (1976) 5188--5192.
\newblock \href {http://dx.doi.org/10.1103/PhysRevB.13.5188}
  {\path{doi:10.1103/PhysRevB.13.5188}}.

\bibitem{Zhao2010}
H.~Zhao, N.~R. Aluru, Temperature and strain-rate dependent fracture strength
  of graphene, J. Appl. Phys. 108~(6) (2010) 064321.
\newblock \href {http://dx.doi.org/10.1063/1.3488620}
  {\path{doi:10.1063/1.3488620}}.

\bibitem{Fasolino2007}
A.~Fasolino, J.~H. Los, M.~I. Katsnelson, Intrinsic ripples in graphene, Nature
  Mater. 6 (2007) 858--861.
\newblock \href {http://dx.doi.org/10.1038/nmat2011}
  {\path{doi:10.1038/nmat2011}}.

\bibitem{Los2005}
J.~H. Los, L.~M. Ghiringhelli, E.~J. Meijer, A.~Fasolino, Improved long-range
  reactive bond-order potential for carbon. {I}. {C}onstruction, Phys. Rev. B
  72 (2005) 214102.
\newblock \href {http://dx.doi.org/10.1103/PhysRevB.72.214102}
  {\path{doi:10.1103/PhysRevB.72.214102}}.

\end{thebibliography}



\end{document}